\documentclass[fleqn,usenatbib]{mnras}

\usepackage{newtxtext,newtxmath}

\usepackage[T1]{fontenc}

\DeclareRobustCommand{\VAN}[3]{#2}
\let\VANthebibliography\thebibliography
\def\thebibliography{\DeclareRobustCommand{\VAN}[3]{##3}\VANthebibliography}

\usepackage{graphicx}	
\usepackage{amsmath}	
\usepackage{threeparttable}
\usepackage{hyperref} 

\title[]{Minute-cadence Observations of the LAMOST Fields with the TMTS:\\ III. Statistic Study of the Flare Stars from the First Two Years}

\author[Qichun Liu et al.]{
Qichun Liu$^{1}$,
Jie Lin$^{1}$\thanks{E-mail:
linjie2019@mail.tsinghua.edu.cn},
Xiaofeng Wang$^{1,2}$\thanks{E-mail:
wang\_xf@mail.tsinghua.edu.cn},
Shenghong Gu$^{3, 4, 6}$,
Jianrong Shi$^{5,6}$,
Liyun Zhang$^{7}$, \newauthor 
Gaobo Xi$^{1}$,
Jun Mo$^{1}$,
Yongzhi Cai$^{3,4,1}$,
Liyang Chen$^{1}$,
Zhihao Chen$^{1}$,
Fangzhou Guo$^{1}$,
Xiaojun Jiang$^{5,6,8}$, \newauthor 
Gaici Li$^{1}$,
Wenxiong Li$^{9}$,
Han Lin$^{1}$,
Weili Lin$^{1}$,
Jialian Liu$^{1}$,
Cheng Miao$^{1}$,
Xiaoran Ma$^{1}$,\newauthor 
Haowei Peng$^{1}$,
Danfeng Xiang$^{1}$,
Shengyu Yan$^{1}$,
Jicheng Zhang$^{10}$,
and Xinhan Zhang$^{1}$
\\
$^{1}$Physics Department and Tsinghua Center for Astrophysics, Tsinghua University, Beijing 100084, China\\
$^{2}$Beijing Planetarium, Beijing Academy of Science and Technology, Beijing 100044, China\\
$^{3}$Yunnan Observatories, Chinese Academy of Sciences, Kunming 650216, China \\
$^{4}$Key Laboratory for the Structure and Evolution of Celestial Objects, Chinese Academy of Sciences, Kunming 650216, China\\
$^{5}$ CAS Key Laboratory of Optical Astronomy, National Astronomical Observatories, Chinese Academy of Sciences, Beijing 100101, People's Republic of China\\ 
$^{6}$ School of Astronomy and Space Science, University of Chinese Academy of Sciences, Beijing 100049, People's Republic of China\\
$^{7}$College of Physics, and College of Big Data and Information Engineering, Guizhou University, Guiyang 550025, China\\
$^{8}$ Center for Astronomical Mega-Science, Chinese Academy of Sciences, 20A Datun Road, Chaoyang District, Beijing 100012, People's Republic of China \\
$^{9}$ The School of Physics and Astronomy, Tel Aviv University, Tel Aviv 69978, Israel \\
$^{10}$ Department of Astronomy, Beijing Normal University, Beijing 100875, People's Republic of China \\}

\date{Accepted XXX. Received YYY; in original form ZZZ}

\pubyear{2015}

\begin{document}
\label{firstpage}
\pagerange{\pageref{firstpage}--\pageref{lastpage}}
\maketitle

\begin{abstract}
Tsinghua University-Ma Huateng Telescopes for Survey (TMTS) aims to detect fast-evolving transients in the Universe, which has led to the discovery of thousands of short-period variables and eclipsing binaries since 2020. In this paper, we present the observed properties of 125 flare stars identified by TMTS within the first two years, with an attempt to constrain their eruption physics. As expected, most of these flares were recorded in late-type red stars with $G_{\rm BP}$ $-$ $G_{\rm RP}$ $> 2.0$ mag; however, the
flares associated with bluer stars tend to be on average more energetic and have broader profiles. The peak flux ($F_{{\rm peak}}$) of the flare is found to depend strongly on the equivalent duration (ED) of the energy release, i.e., $F_{{\rm peak}} \propto {\rm ED}^{0.72\pm0.04}$, which is
consistent with results derived from the Kepler and Evryscope samples. This relation is likely related to the magnetic loop emission, while, for the more popular non-thermal electron heating model, a specific time evolution may be required to generate this relation. We notice that flares produced by hotter stars have a flatter $F_{{\rm peak}} \propto {\rm ED}$ relation compared to that from cooler stars.
This is related to the statistical discrepancy in light-curve shape of flare events with different colors. In spectra from LAMOST, we find that flare stars have apparently stronger H$\alpha$ emission than inactive stars, especially at the low temperature end, suggesting that chromospheric activity plays an important role in producing flares. On the other hand, the subclass with frequent flares is found to show H$\alpha$ emission of similar strength in its spectra to that recorded with only a single flare but similar effective temperature, implying that the chromospheric activity may not be the only trigger for eruptions.

\end{abstract}




\section{Introduction}
\label{introduction}

Stellar flare represents outburst of stellar atmosphere, accompanied by a sudden release of radiative energy over a wide range of electromagnetic wave ~\citep{Kowalski_2013, 2020ApJ...902..115H}. The convective motion in the stellar photosphere causes the magnetic fields to reconnect, which converts the stored energy into kinetic energy and thermal energy~\citep{Allred2015}. The released energy can heat the reconnection region, accelerating and propagating the charged particles along the magnetic flux loop. These particles then collide and heat the plasma to a temperature of $10^6$ $-$ $10^7$ ${\rm K}$, resulting in significant flux enhancement in a short time. This phenomenon has been extensively studied since the Carrington event was observed on the sun~\citep{1859MNRAS..20...13C}. However, the exact physics underlying stellar flares is still controversial. 

White-light flares (WLFs) are those flares detected in visible light, which are thought to have physical relations with flare emissions in hard X-rays, suggesting that high-energy electrons are essential to the outburst in white light~\citep{2011ApJ...739...96K, 2018IAUS..340..221N}. For example, \citet{1970SoPh...15..176N} suggested that heating of photosphere by energetic particles would give rise to WLFs in stellar atmosphere, while \citet{1989SoPh..124..303M} proposed that chromospheric activity can provide sufficient energy for sudden increase in photospheric radiation. With the help of radiative hydrodynamic method, the WLFs can be well simulated ~\citep{2005ApJ...630..573A, 2019ApJ...882...97P}. The flare emission usually lasts for 10-20 minutes, with the light curves being characterized by a rapid rise and an exponential decay (FRED) phase~\citep{2014ApJ...797..122D}. The chromospheric activity is believed to be critical for the production of flare emission, and the spectral lines like Ca H\&K and Blamer lines are used as indicators to diagnose the physical process underlying the  outbursts~\citep{2021ApJS..253...51L, 2022ApJ...928..180W}. 

High-cadence surveys, such as Kepler Space Telescope~\citep{2010Sci...327..977B}, the Transiting Exoplanet Survey Satellite~\citep{10.1117/1.JATIS.1.1.014003}, Evryscope~\citep{2019ApJ...881....9H}, the Next Generation Transit Survey (NGTS)~\citep{2021MNRAS.504.3246J} discovered numerous WFLs with well-sampled light curves, pushing forward the understanding of correlations between flares and stellar properties such as rotation, color, magnetic activity, and age etc. Statistically, the fraction of producing flares is found to increase significantly for stars with redder colors, i.e., M dwarfs, while flare activity is thought to decrease with age of the stars~\citep{2009AJ....138..633K, 2019A&A...622A.133I}. Previous studies revealed that there are a sufficient amount of earth-like planets orbiting the M dwarfs, so the activities of M dwarfs are important for the study of habitability of planets~\citep{2015ApJ...807...45D, 2019ApJ...881....9H}, as the planets in habitable zones are usually very close to the host stars. The stellar flares can affect the planetary atmospheres and cause transmission with bio-indicating chemical species~\citep{2021NatAs...5..298C}.

During the minute-cadence monitoring of the LAMOST fields since 2020, the TMTS also recorded numerous flare events besides the short-period pulsating stars and eclipsing binaries~\citep{2022MNRAS.509.2362L}. In this work, we present the photometric and spectroscopic properties of the flare events detected by the TMTS. Our study benefits from updated distances from the Gaia DR2 database~\citep{2018A&A...616A...1G}, and spectral data provided by the LAMOST \citep{2015RAA....15.1095L} 
can further help constrain the physical origin of stellar flares. The Gaia DR2- and LAMOST-TMTS cross matching enable statistical studies of the correlations between chromosheric activity and observed flares in different types of stars. In section ~\ref{ob}, we describe our TMTS flare observations, the method for the analysis of the TMTS light curves, and the LAMOST spectra. The results of our work are presented in section ~\ref{re}, and relative discussion is given in section ~\ref{discussion}. We summarize our results in section ~\ref{summary}.

\section{Observations and Data Analysis}
\label{ob}

\subsection{TMTS light curves}
\label{lightcurve}
Since the beginning of 2020, TMTS started to monitor the LAMOST fields with a cadence of about 1 minute and durations ranging from about 2.5-12 h. The time intervals of each data point are in the range of 60-120 s, and most of them (i.e., 88\%) are below 90 s, including the readout time. Luminous filters are used in the TMTS observations, which cover the wavelength from 330 nm to about 900 nm. With the QHY 4040 CMOS detector, the TMTS system can detect objects down to $\sim$19.0 mag with a pixel scale of  $1.^{\prime \prime}86$ (see more details in ~\citet{2020PASP..132l5001Z}). 

During the first two-year survey, the TMTS has produced millions of uninterrupted light curves and has revealed a series of interesting short-period variable stars (e.g. \citealt{2022NatAs.tmp..217L}) with minute-cadence observations. We select the flare candidates based on the two-year observation data, which contains 10,856,233 uninterrupted light curves with at least 100 valid measurements for about ten million sources, and the method for flare search has been described in ~\citet{2022MNRAS.509.2362L}. The flare phenomenon are identified by examining the outlier features in the light curves, while non-flare variations are removed by fitting a composite model of 4th-order Fourier series. In our MCMC fitting procedure, we added a systematic uncertainty $\sigma_s$ to the data likelihood; for each data point, we calculate the error $\sigma=\sqrt{{\sigma_s}^2 + {\epsilon}^2}$, where $\epsilon$ is the measurement error of each point. Outlier photometric data points with confidence level $<$ 3~$\sigma$ and without showing clear profiles, were not identified as flares. The flare candidates were further checked by light curve profiles and raw images. Finally, a total of 125 sources were defined as flare stars with at least one flare recorded during our observations. Furthermore, 46 stars have spectra and atmospheric parameters from the LAMOST DR7. The details of these 125 flares are listed in Table~\ref{tab1}. 

\begin{figure}
	\includegraphics[width=\columnwidth]{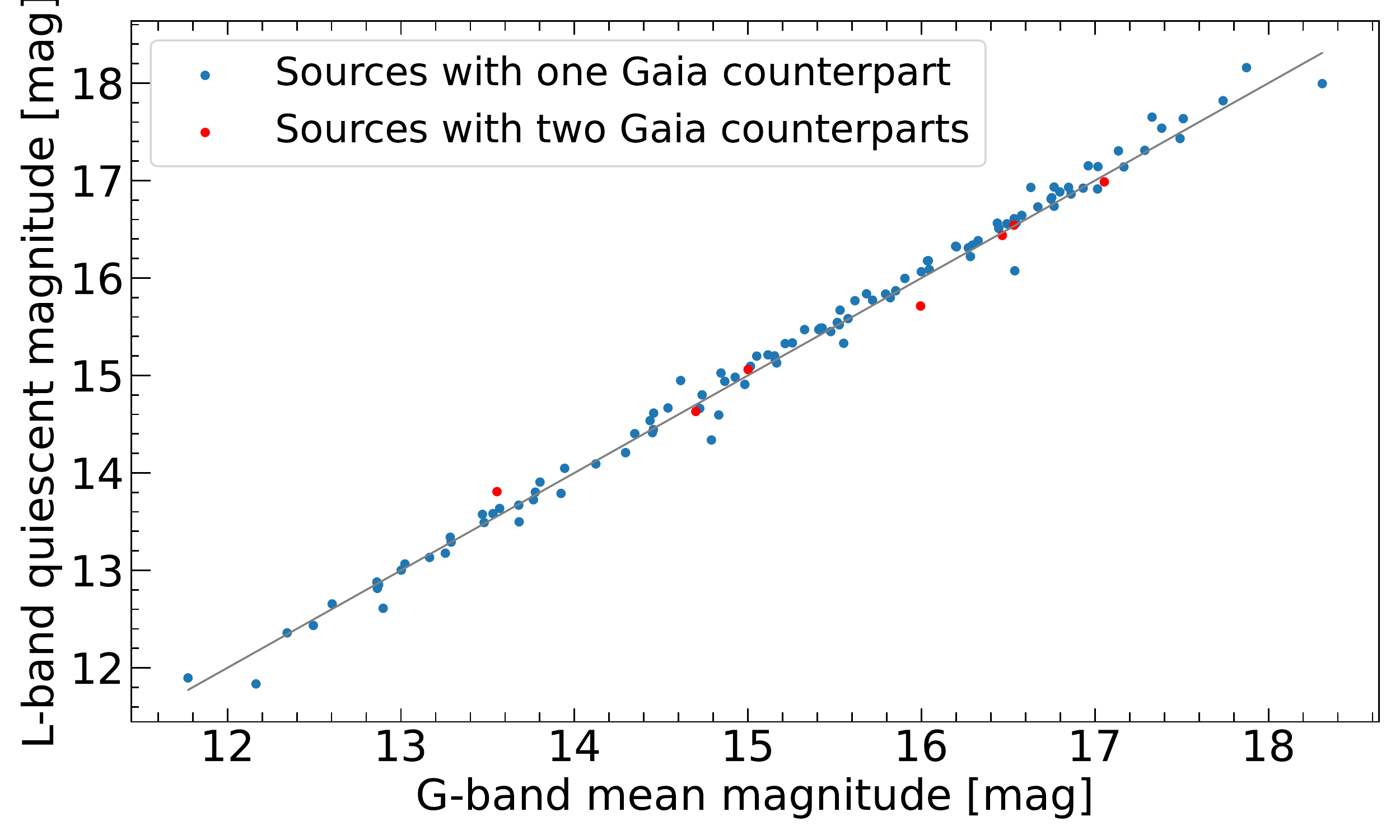}
    \caption{The comparison between G-band mean magnitude and L-band magnitude for the sources with selected Gaia counterparts. The blue points indicate sources with only one counterpart in the cross-matching, while red points indicate sources with two counterparts. The G-band magnitudes for red points are from the final picked counterparts. The gray dashed line represents the diagonal line.}
    \label{magnitude_compare}
\end{figure}

Considering the pixel resolution of TMTS, we cross-matched our sample with Gaia sources within a $3.^{\prime \prime}0$ aperture. Finally, 108 flare stars are found to have only one Gaia counterpart, while the rest 17 ones have two Gaia counterparts. 
For each of these 17 stars, we select the closer one of the two counterparts relative to the TMTS positions, which results in 7 ones having G-band magnitudes brighter than the other counterparts by $2~{\rm mag}$. For the remained 10 sources with relatively fainter, closer counterparts, we only included them in the relation of $F_{\rm peak}-{\rm ED}$ and the timescale analysis in section~\ref{Timescale}, but not in the statistics of Gaia parameters. Analysis by ~\citet{2022MNRAS.509.2362L} already showed that the TMTS $L$-band magnitudes are basically consistent with the Gaia G-band magnitudes. We also used the quiescent-state flux $f_0$ derived from the fitting described below (see Eq.~\ref{eqmodel}) to get the quiescent magnitudes, and compared them with the G-band mean magnitudes of the selected Gaia counterparts. As shown in Fig.~\ref{magnitude_compare}, these two magnitudes are well consistent with each other for the cases with one or two Gaia counterparts. Parameters of these 10 flare stars from the TMTS are also listed in Table~\ref{tab1} and Table~\ref{tab2}. In the end, 93 stars have reliable Gaia parallax measurements ($\sigma_{\varpi}/{\varpi}\leq 0.2$) and Gaia colors with reddening-removed ($G_{\rm BP}-G_{\rm RP}$) information from the Gaia-DR2 database. Spectral parameters of 40 stars were used in this work.

\begin{figure*}
	
	\includegraphics[width=2\columnwidth]{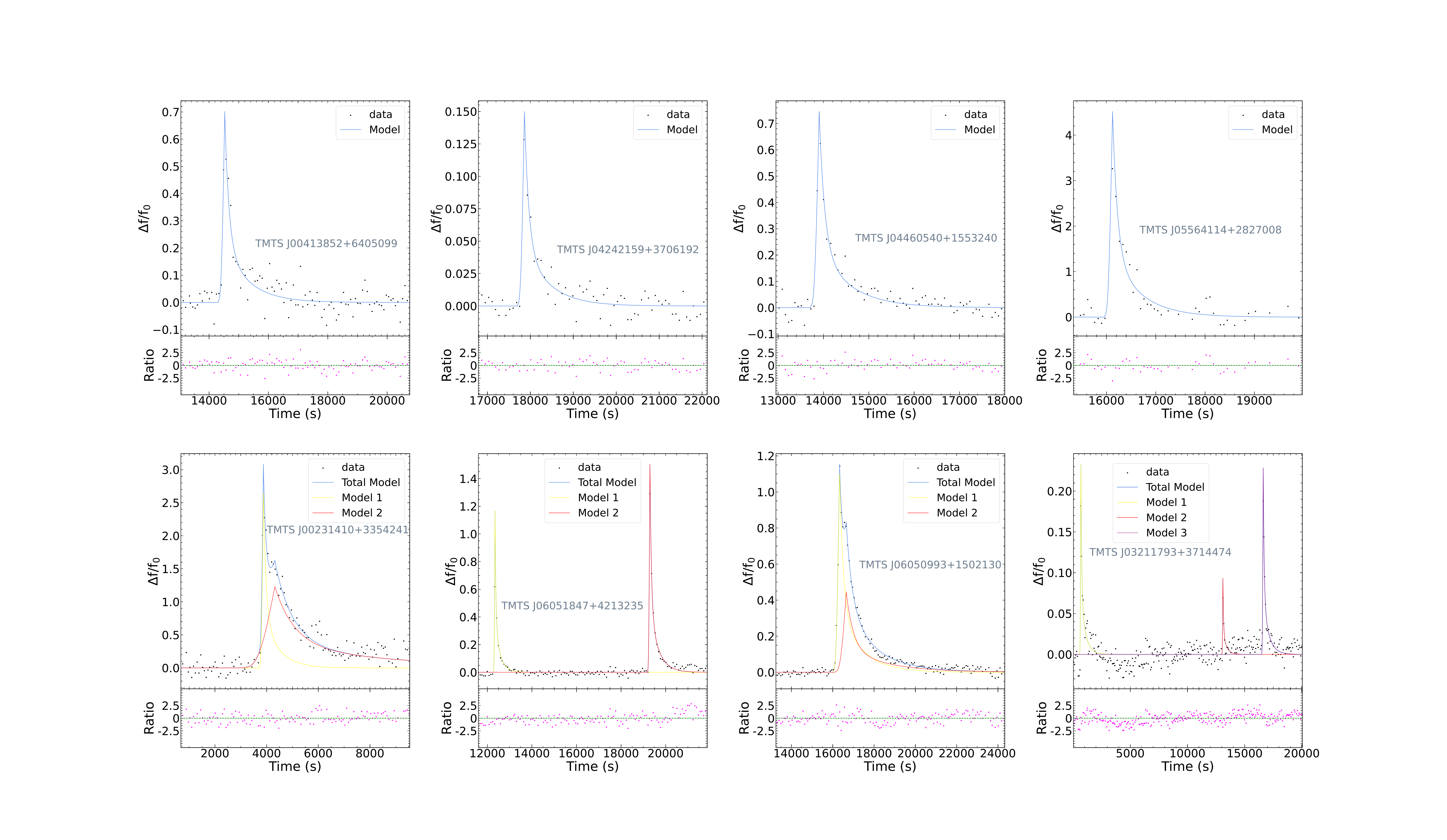}
	\caption{Examples of the observed flare light curves and the best-fit models. The upper panels show the classical flares, while the lower panels represent the complex events. In each individual plot of the flare light curve, the bottom part represents the residual obtained by subtracting the best-fit model from the observed data.}
    \label{Fig1}
\end{figure*}

In analysis of the flare light curves, we adopted the empirical template constructed by ~\citet{2014ApJ...797..122D}. This template has two free parameters, peak flux $f_{\rm peak}$ and the full-time width at half maximum flux $t_{1/2}$. Following ~\citet{2014ApJ...797..122D}, $t_{1/2}$ is also set to be equal to the rise time of flare events, namely $t_{\rm peak}-t_0$ (See Eq.~\ref{eqmodel} below), where $t_{\rm peak}$ and $t_0$ represent the peak and start time of the flares, respectively. For the analysis of our sample, this specific model also works well because most data points have confidence levels < 2.5 $\sigma$ and we do not have any unusual flares in our sample. To directly fit the flare light curves, we modify this flare template and set four parameters, including the peak flux $f_{\rm peak}$, the time of flare peak $t_{\rm peak}$, the start time $t_0$, and the quiescent-state flux $f_0$. We set $t'=\frac{t-t_{\rm peak}}{t_{\rm peak}-t_0}$, then the function of $t$ for the template can be written as 

\begin{equation}
\label{eqmodel}
f(t)=\left\{
\begin{tiny}\begin{aligned}
f_0 & , & t'<-1, \\
A(1+1.941\ t'-0.175\ t'^2-2.246\ t'^3-1.125\ t'^4) + f_0 & , & -1\leq t'<0, \\
A(0.6890\ e^{-1.6t'}+0.3030\ e^{-0.2783t'}) + f_0 & , & t'\geq 0.
\end{aligned}\end{tiny}
\right.
\end{equation}
Where $A$ is set as $f_{\rm peak}-f_0$, and more discussions of this model are given in section~\ref{re}. We ran the Markov chain Monte Carlo (MCMC) method to fit the data. In order to get a reasonable prior guess, we first linearly interpolated the light curve to have a 0.1-minute resolution, and then we set the median value of the flux ($f_{\rm 0, obs}$) as a guessed flux in quiescent state. The observed amplitude $f_{\rm peak, obs}$ is the maximum value of the interpolated data, and $t_{\rm peak, obs}$ is the corresponding time of peak light. The $t_{\rm FWHM,obs}$, defined as the interval between the first and last data point with flux larger than $(f_{\rm 0, obs}+f_{\rm peak, obs})/2$, was then measured for each light curve. The measured $t_{\rm peak,obs}-t_{\rm FWHM, obs}$ is set as the initial guess of the start time. To avoid possible effects of light curve fluctuations caused by atmospheric extinction, the data selected for the fit are in the range $t_{\rm peak,obs}-5~t_{\rm FWHM,obs}\leq t\leq t_{\rm peak,obs}+20~t_{\rm FWHM,obs}$. Note that the interpolated data were only used for determining the prior guesses and this range, while the data in the fitting procedure were un-interpolated light curves. According to \citet{2014ApJ...797..122D} and \citet{2020ApJ...902..115H}, we fit the light curves of complex events with the combination of two or three classical flare models. The complex events include several single flare events occurring at different epochs without overlapping (multi-flare) and an event with multiple peaks (multi-peak). In the following analysis, we considered a multi-flare event as several single flare events and a multi-peak event as an individual event. For the fit of two-flare composite model, if the fractional flux (${\Delta f/f_0}$, which will be mentioned in the paragraph below) of the first flare at the beginning of the second flare is less than 5 percent, we identified it as a multi-flare event. The same strategy is also applied to three-flare model fits. Since it is difficult to define a global timescale for multi-peak events and $t_{1/2}$ is only valid in a single component, we did not include these events in the analysis of $t_{1/2}$ below. The duration time of each flare event of our sample is listed in Table \ref{tab2}. The light curves and the corresponding fitting results are shown in Fig.~\ref{Fig1}.

\begin{table*}
    \scriptsize
    \centering
	\caption{Summary of properties of 125 flare stars from the first two-year survey of TMTS. }
	\label{tab1}
	\begin{threeparttable}
	\begin{tabular}{llllllllll} 
		\hline
		Name & Spt & Parallax & Distance  & Abs.Mag.G & $G_{\rm BP}-G_{\rm RP}$ & $T_{\rm eff}$ & log $g$ & ${\rm Fe/H}$ & $ {\rm EW}_{\rm H \alpha}$\\
		 & & mas & kpc & & & K & & &\AA\\
		\hline
		TMTS J00182181+6418398& K&$4.75 \pm 0.02$&$0.2106 \pm 0.0008$&$6.923 \pm 0.009$&$1.456 \pm 0.009$&&&&\\
TMTS J00231410+3354241& M (dM5)&$9.07 \pm 0.13$&$0.110 \pm 0.002$&$10.83 \pm 0.03$&$2.906 \pm 0.013$&$3296.53 \pm 85.77$&$5.25 \pm 0.14$&&$9.5 \pm 0.5$\\
TMTS J00295311+3351573& M&$7.4 \pm 0.2$&$0.136 \pm 0.003$&$11.39 \pm 0.05$&$3.07 \pm 0.03$&&&&\\
TMTS J00413852+6405099& M&$16.61 \pm 0.04$&$0.0602 \pm 0.0002$&$10.947 \pm 0.006$&$2.864 \pm 0.006$&&&&\\
TMTS J01022450+6103522& M&$9.4 \pm 0.4$&$0.107 \pm 0.005$&$7.02 \pm 0.10$&$1.879 \pm 0.004$&&&&\\
TMTS J01104648+6511446& M&$3.97 \pm 0.07$&$0.252 \pm 0.005$&$9.72 \pm 0.04$&$2.547 \pm 0.014$&&&&\\
TMTS J01294586+7622160& M&$6.4 \pm 0.2$&$0.157 \pm 0.005$&$9.55 \pm 0.07$&$2.450 \pm 0.005$&&&&\\
TMTS J01311947+4608201& M&$4.20 \pm 0.12$&$0.238 \pm 0.007$&$8.20 \pm 0.06$&$2.260 \pm 0.007$&&&&\\
TMTS J01344581+5559022& K&$1.42 \pm 0.08$&$0.70 \pm 0.04$&$6.99 \pm 0.12$&$1.632 \pm 0.009$&&&&\\
TMTS J01355434+5428597& M&$3.80 \pm 0.11$&$0.263 \pm 0.007$&$10.18 \pm 0.06$&$2.83 \pm 0.02$&&&&\\
TMTS J01380974+4723211& &$0.44 \pm 0.13$&&&&&&&\\
TMTS J01442812+4327127& M&$7.38 \pm 0.08$&$0.1354 \pm 0.0014$&$9.92 \pm 0.02$&$2.807 \pm 0.014$&&&&\\
TMTS J02012142+7430235& M&$5.39 \pm 0.08$&$0.185 \pm 0.003$&$10.62 \pm 0.03$&$2.69 \pm 0.02$&&&&\\
TMTS J02093570+4117071& M&$3.42 \pm 0.06$&$0.292 \pm 0.005$&$7.89 \pm 0.04$&$2.082 \pm 0.006$&&&&\\
TMTS J02354528+3429199& M (dM5)&$32.06 \pm 0.09$&$0.03 \pm 0.00$&$11.473 \pm 0.006$&$2.957 \pm 0.006$&$3284.97 \pm 61.49$&$5.34 \pm 0.10$&&$4.4 \pm 0.4$\\
TMTS J02382397+4342369& M&$3.07 \pm 0.07$&$0.325 \pm 0.007$&$7.19 \pm 0.05$&$1.888 \pm 0.005$&&&&\\
TMTS J02420089+3852565& K&$1.31 \pm 0.11$&$0.76 \pm 0.07$&$7.0 \pm 0.2$&$1.59 \pm 0.02$&&&&\\
TMTS J02461232+3409287& K&$0.91 \pm 0.05$&$1.10 \pm 0.06$&$5.40 \pm 0.12$&$1.022 \pm 0.006$&&&&\\
TMTS J02515141+4434084& &&&&&&&&\\
TMTS J02552795+5815041& M&$9.4 \pm 0.2$&$0.106 \pm 0.002$&$8.40 \pm 0.05$&$2.177 \pm 0.007$&&&&\\
TMTS J02570023+4642138& M&$14.43 \pm 0.10$&$0.0693 \pm 0.0005$&$10.34 \pm 0.02$&$2.966 \pm 0.008$&&&&\\
TMTS J03060058+2807084& &&&&&$3536.59 \pm 57.14$&$5.08 \pm 0.09$&&$3.4 \pm 0.2$\\
TMTS J03070420+0352598& M (dM4)&$18.13 \pm 1.29$&$0.055 \pm 0.004$&$8.9 \pm 0.2$&$2.422 \pm 0.006$&$3623.81 \pm 83.70$&$5.28 \pm 0.14$&&$2.35 \pm 0.12$\\
TMTS J03100245+4346147& M&$2.28 \pm 0.11$&$0.44 \pm 0.02$&$8.63 \pm 0.11$&$2.09 \pm 0.02$&&&&\\
TMTS J03111364+4526272& &&&&&$3458.15 \pm 88.03$&$5.00 \pm 0.15$&&$5.7 \pm 0.3$\\
TMTS J03143182+5700585& M&$8.21 \pm 0.07$&$0.1218 \pm 0.0010$&$10.09 \pm 0.02$&$2.862 \pm 0.012$&&&&\\
TMTS J03211793+3714474& M (dM4)&$15.42 \pm 0.13$&$0.0649 \pm 0.0006$&$10.29 \pm 0.02$&$2.621 \pm 0.010$&$3469.99 \pm 47.17$&$5.48 \pm 0.08$&&$4.3 \pm 0.3$\\
TMTS J03232062+3459324$^{\star}$& &$0.22 \pm 0.09$&&&&&&&\\
TMTS J03264186+3253138& K (dM0)&$7.0 \pm 0.3$&$0.144 \pm 0.006$&$7.50 \pm 0.09$&$1.759 \pm 0.007$&$3984.74 \pm 95.67$&$4.8 \pm 0.2$&$-0.48 \pm 0.13$&$0.51 \pm 0.09$\\
TMTS J03272410+7459407& K&$6.74 \pm 0.10$&$0.148 \pm 0.002$&$7.17 \pm 0.03$&$1.715 \pm 0.007$&&&&\\
TMTS J03324904+5755181& &&&&&&&&\\
TMTS J03392990+4117185& &&&&&$3655.91 \pm 154.51$&$5.3 \pm 0.2$&&$1.6 \pm 0.2$\\
TMTS J03402568+3501575& K&$2.5 \pm 0.3$&$0.40 \pm 0.04$&$6.5 \pm 0.2$&$1.664 \pm 0.004$&&&&\\
TMTS J03423772+3805512& &$0.22 \pm 0.07$&&&&&&&\\
TMTS J03474651+3635558& M (dM3)&$5.74 \pm 0.09$&$0.174 \pm 0.003$&$9.79 \pm 0.03$&$2.887 \pm 0.014$&$3338.56 \pm 83.22$&$5.16 \pm 0.14$&&$5.4 \pm 0.4$\\
TMTS J03475843+4142053& M&$8.46 \pm 0.08$&$0.1182 \pm 0.0011$&$10.04 \pm 0.02$&$2.801 \pm 0.009$&&&&\\
TMTS J03525194+5121166& M&$11.33 \pm 0.06$&$0.0882 \pm 0.0005$&$11.092 \pm 0.012$&$2.840 \pm 0.014$&&&&\\
TMTS J03581344+5404505& M&$6.74 \pm 0.04$&$0.1484 \pm 0.0010$&$9.157 \pm 0.014$&$2.560 \pm 0.004$&&&&\\
TMTS J04115590+1722272& K (K5)&$7.70 \pm 0.04$&$0.1299 \pm 0.0007$&$7.295 \pm 0.013$&$1.548 \pm 0.010$&$4312.63 \pm 212.07$&$4.3 \pm 0.3$&$-0.1 \pm 0.2$&$1.4 \pm 0.2$\\
TMTS J04150931+1904271& &&&&&&&&$7.1 \pm 0.4$\\
TMTS J04192986+2145136& M (dM4)&$20.90 \pm 0.07$&$0.0478 \pm 0.0002$&$9.463 \pm 0.008$&$2.547 \pm 0.007$&&&&$4.7 \pm 0.3$\\
TMTS J04222375+4137066$^{\star}$& K&$2.02 \pm 0.05$&$0.496 \pm 0.013$&$6.68 \pm 0.06$&$1.309 \pm 0.004$&&&&\\
TMTS J04242159+3706192& M (dM5)&$19.26 \pm 0.08$&$0.0519 \pm 0.0002$&$9.712 \pm 0.009$&$2.942 \pm 0.007$&$3202.26 \pm 46.59$&$4.40 \pm 0.08$&&$7.8 \pm 0.3$\\
TMTS J04254360+3501496& K&$5.14 \pm 0.03$&$0.1946 \pm 0.0011$&$6.381 \pm 0.012$&$1.388 \pm 0.009$&&&&\\
TMTS J04282995+3747418& M (dM1)&$3.92 \pm 0.09$&$0.255 \pm 0.006$&$8.79 \pm 0.05$&$2.17 \pm 0.02$&$3682.89 \pm 56.03$&$5.45 \pm 0.09$&&$3.4 \pm 0.2$\\
TMTS J04291264+2154545$^{\star}$& &$22.5 \pm 0.2$&$0.0445 \pm 0.0004$&$11.12 \pm 0.02$&&$3435.17 \pm 52.21$&$5.27 \pm 0.09$&&$1.00 \pm 0.15$\\
TMTS J04460540+1553240& M&$18.17 \pm 0.08$&$0.0550 \pm 0.0002$&$11.412 \pm 0.010$&$3.107 \pm 0.007$&&&&\\
TMTS J04514921+4341465& M&$4.81 \pm 0.08$&$0.208 \pm 0.004$&$8.40 \pm 0.04$&$2.419 \pm 0.008$&&&&\\
TMTS J04541626+3459537& M&$3.6 \pm 0.2$&$0.278 \pm 0.014$&$9.39 \pm 0.11$&$2.51 \pm 0.04$&&&&\\
TMTS J04563049+4348567& M (dM2)&$7.47 \pm 0.06$&$0.1339 \pm 0.0011$&$8.82 \pm 0.02$&$2.431 \pm 0.005$&$3656.39 \pm 76.65$&$5.50 \pm 0.13$&&$3.6 \pm 0.6$\\

		\hline
	\end{tabular}
	\begin{tablenotes}[normal, flushleft]
	        \item 
	        \textit{Note}: Parallax, distance, G band absolute magnitude, Gaia Bp-Rp are from Gaia DR2. $T_{\rm eff}$, log $g$, Fe/H are from LAMOST DR7. The spectral types are estimated from Gaia color based on the updated table\textsuperscript{~\ref{spectral_estimate}} in ~\citet{2013ApJS..208....9P}, and the spectral types from LAMOST DR7 are given in the brackets. ${\rm EW}_{\rm H \alpha}$ is the equivalent width of ${\rm H \alpha}$ line measured in section ~\ref{LAMOST}. $\star$ means the stars have unreliable Gaia cross-matched sources. This table is only a subset and the full table is in machine-readable form and is available online.
	\end{tablenotes}
	\end{threeparttable}
\end{table*}

With the flare model described above, we derived the physical parameters for each of our flare sample, such as  $t_{1/2}$ defined above and the relative (or fractional) flux defined by $\Delta f/f_0=\frac{f-f_0}{f_0}$, where $f_0$ represents the quiescent flux from the model. In the following text, we denoted $(\Delta f/f_0)_{\rm peak}$ as $F_{\rm peak}$ for convenience. The equivalent duration ~\citep[ED, see ][]{1972Ap&SS..19...75G} is the area under the fractional flux of the light curve, as described by ~\citet{2014ApJ...797..121H}. We calculated ED by integrating our model fits. The physical meaning of the ED measured for a flare star is the time that it takes to release the same amount of flare energy in its quiescent state. In order to calculate the flare energy, we need to estimate the quiescent-state luminosity $L_0$. Since we already know the quiescent flux $f_0$, the white-light quiescent magnitude $m_{\rm L}$ can be derived from the relation below \begin{equation}
    m_{\rm L}=-2.5~{\rm log_{10}}~ f_{\rm 0}+zp,
	\label{eq1}
\end{equation} 
where $zp$ is the photometric zeropoint~\citep{2018A&A...616A...8A}, and it is determined from each TMTS observation. By using the flux calibration of $m_{\rm L}=0$, we got intrinsic fluxes of these stars. These fluxes were multiplied by $4\pi d^2$, where $d$ is the distance from Gaia DR2, then we got the quiescent luminosity $L_0$ for our sample. The energy released in white light or $L$ band is calculated by $E_{\rm L}={\rm ED} \times L_0$, and the error is calculated by error propagation. For the convenience of comparing the $L$-band energy with other work, we converted $E_{\rm L}$ into bolometric energy $E_{\rm bol}$ by assuming that the flare energy distribution can be roughly approximated as a blackbody of 10,000 K ~\citep[see, eg.,][]{2017ApJ...838...22G, 2019ApJ...876..115S}. We convoluted the 10,000 K blackbody with the transmission curve of Luminous filter and the quantum efficiency of the QHY 4040 CMOS detector, yielding $E_{\rm L}=0.29\ E_{\rm bol}$. The fitting results are listed in Table~\ref{tab2}. The uncertainties reported in Table~\ref{tab2} were calculated by the MCMC samples except for $E_{\rm L}$ and $E_{\rm bol}$.

\subsection{Spectra}
The spectra of our sample are obtained by crosschecking with the database of the Large Sky Area Multi-Object Fiber Spectroscopic Telescope (LAMOST; ~\citet{2012RAA....12.1197C}), which is also located at Xinglong Station of NAOC. The LAMOST has an effective aperture of 4 meters and a field of view (FOV) of $5^{\circ}$, which operates in two observation modes: low-resolution (LR) mode and medium-resolution (MR) mode, with R=1800 and R=7500, respectively~\citep{2021ApJS..253...51L}. The regular LR survey of LAMOST was launched since Oct. 2012, resulting in more than 10 million spectra up to 2018~\citep{2015RAA....15.1089L, 2020arXiv200507210L}. Since 2018, LAMOST started the medium-resolution spectroscopic survey.

\begin{figure*}
	\includegraphics[width=2\columnwidth]{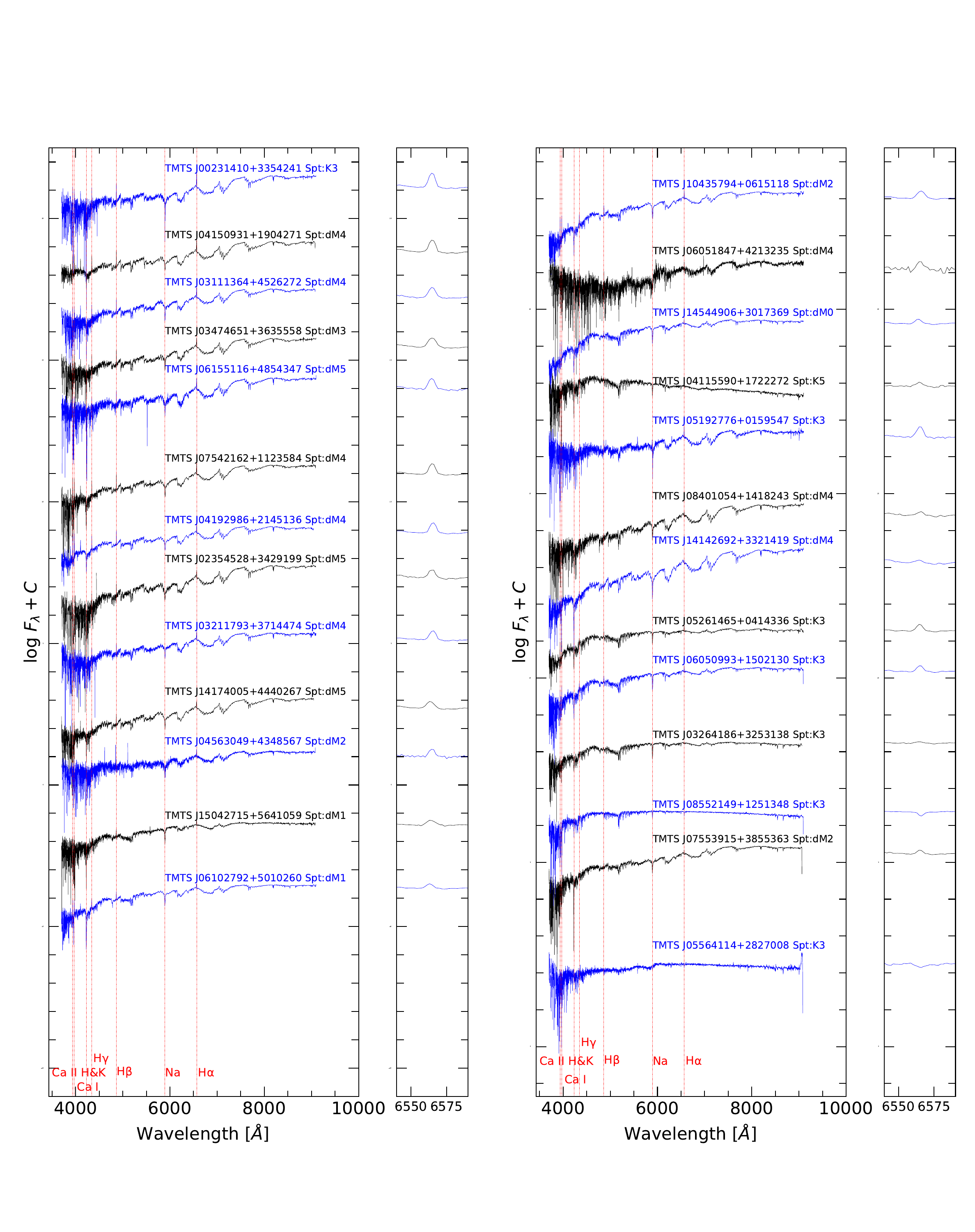}
    \caption{The LAMOST low-resolution spectra for part of our flare sample with single-epoch spectroscopic observation. The right panel of each plot shows the zoomed ${\rm H \alpha}$ feature for each spectrum. The blue and black color are used to distinguish adjacent spectra.}
    \label{lamost_spectral}
\end{figure*}

\begin{figure*}
	\includegraphics[width=2\columnwidth]{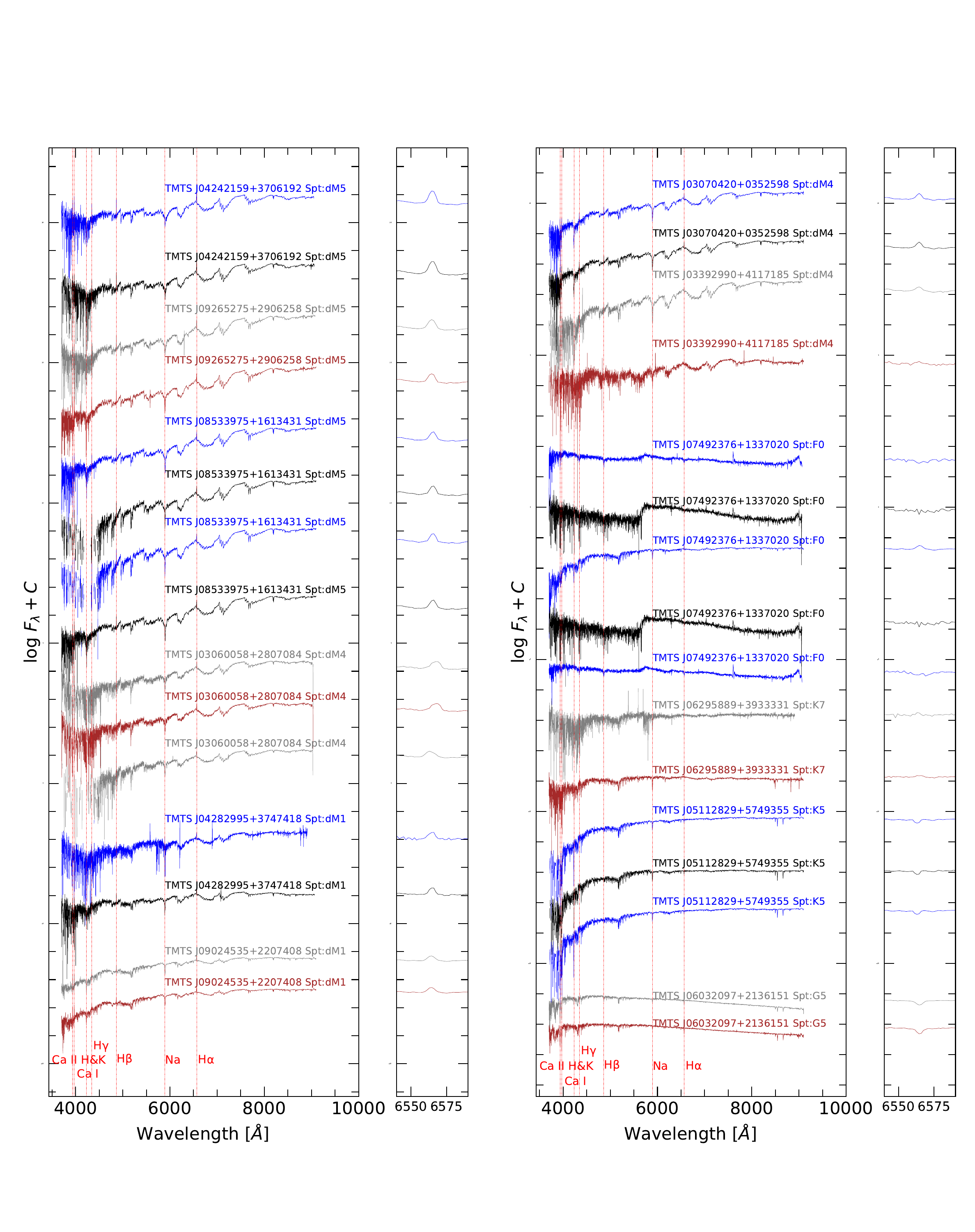}
    \caption{The LAMOST low-resolution spectra for part of our flare sample with multiple spectroscopic observations. The right panel of each plot shows the zoomed ${\rm H \alpha}$ feature for each spectrum. The bule and black (or grey and brown) color are used to distinguish adjacent spectra of each source.}
    \label{lamost_spectral_multi_time}
\end{figure*}

 A total of 40 flare stars of our TMTS sample are found to have LAMOST spectra, including 57 low resolution spectra and 2 median resolution spectra. The spectra of these flare sample are shown in Fig.~\ref{lamost_spectral} and Fig.~\ref{lamost_spectral_multi_time}, respectively, where the ${\rm H \alpha}$ lines are zoomed in the right panel of the plots. One can see that most of these spectra are characterized by $\rm Ca ~\uppercase\expandafter{\romannumeral2} ~H\&K$ lines, Balmer emission lines,  ${\rm Na ~\lambda5890}$ absorption line, and ${\rm Ca ~\uppercase\expandafter{\romannumeral1}} {~\lambda 4227}$ absorption line. 
Among them, the $\rm Ca ~\uppercase\expandafter{\romannumeral2} ~H\&K$  and Balmer lines are conventional indicators of chromospheric activity ~\citep{Mamajek_2008, 2014ApJ...795..161D, Kiman_2019}, and their strength can be used to quantitatively examine the correlation between properties of flares and chromosperic activity. 
For the sources with multiple observations, we examined the main spectral features and did not find any significant variations (see also Fig.~\ref{lamost_spectral_multi_time}). 

We further measured the equivalent width (EW) of ${\rm H \alpha}$ line (${\rm EW}_{\rm H \alpha}$) following the method described by ~\citet{10.1093/mnras/stw1923}. The continuum flux is taken as the average value in the range 6547-6557\ \AA\ and 6570-6580 \AA, respectively, and the error is the standard deviation of the flux in these two measurement regions. The ${\rm EW}_{\rm H \alpha}$ is then calculated by integrating the flux in the wavelength range 6557-6569 \AA. In the measurement of ${\rm H \alpha}$ emission, we only include those spectra with signal to noise ratio (SNR) $>$5 \footnote{Note that the SNR of the spectra is estimated by ${\rm flux}\times  \sqrt{\rm inverse~variance}$}. The error of ${\rm EW}_{\rm H \alpha}$ is calculated by error propagation with the flux uncertainty per pixel in the integrated band and the error of continuum flux. For sources with multiple observations, the error-weighted mean value of ${\rm H \alpha}$ emission was taken as the corresponding ${\rm EW}_{\rm H \alpha}$ of the source. All the measured ${\rm EW}_{\rm H \alpha}$ are given in Table ~\ref{tab1}.   

\begin{table*}
    \scriptsize
    \centering
	\caption{Summary of the flare parameters for 125 flare stars from the first two-year survey of TMTS. }
	\label{tab2}
	\begin{threeparttable}

	\begin{tabular}{lllllllll} 
		\hline
		Name & Start Time & Peak Flux& $t_0$ & $t_{\rm 1/2}$  & Duration & ED & log $E_{\rm L}$ & log $E_{\rm bol}$ \\
		 & MJD& ${\Delta f}/f$ &s  & s & s & s & log [erg] & log [erg]\\
		\hline
		TMTS J00182181+6418398& 59507.44994&$0.43 \pm 0.04$&$33683.89 \pm 74.41$&$512.78 \pm 83.00$&$3802.80 \pm 613.20$&$399.15 \pm 57.83$&$34.82 \pm 0.06$&$35.36 \pm 0.06$\\
TMTS J00295311+3351573& 59148.44718&$2.77 \pm 1.19$&$987.12 \pm 69.25$&$207.45 \pm 76.35$&$2105.48 \pm 429.27$&$912.31 \pm 109.76$&$33.52 \pm 0.06$&$34.06 \pm 0.06$\\
TMTS J00413852+6405099& 59507.44985&$0.70 \pm 0.05$&$14306.08 \pm 24.58$&$222.30 \pm 26.33$&$1648.11 \pm 195.17$&$284.06 \pm 23.96$&$33.10 \pm 0.04$&$33.64 \pm 0.04$\\
TMTS J01022450+6103522& 59509.4611&$0.128 \pm 0.011$&$3553.72 \pm 13.21$&$142.36 \pm 17.20$&$1055.43 \pm 127.49$&$32.90 \pm 1.96$&$33.93 \pm 0.05$&$34.47 \pm 0.05$\\
TMTS J01104648+6511446& 59177.41106&$2.6 \pm 0.5$&$946.68 \pm 22.74$&$129.27 \pm 26.08$&$1260.79 \pm 292.26$&$599.52 \pm 48.10$&$33.98 \pm 0.04$&$34.52 \pm 0.04$\\
TMTS J01294586+7622160& 59185.47665&$1.99 \pm 0.07$&$8031.89 \pm 11.89$&$197.55 \pm 11.58$&$1464.62 \pm 85.85$&$715.57 \pm 33.00$&$34.07 \pm 0.04$&$34.61 \pm 0.04$\\
TMTS J01311947+4608201& 59510.46878&$0.43 \pm 0.02$&$24273.21 \pm 67.83$&$1047.73 \pm 74.47$&$7767.82 \pm 552.13$&$825.21 \pm 49.76$&$34.63 \pm 0.04$&$35.17 \pm 0.04$\\
TMTS J01344581+5559022& 59151.45242&$1.01 \pm 0.07$&$21908.77 \pm 59.95$&$601.82 \pm 66.44$&$4462.47 \pm 492.42$&$1101.87 \pm 89.80$&$35.09 \pm 0.06$&$35.63 \pm 0.06$\\
TMTS J01355434+5428597& 59151.45571&$17.45 \pm 4.15$&$31062.65 \pm 11.71$&$74.38 \pm 12.72$&$553.87 \pm 94.50$&$2286.92 \pm 286.04$&$34.37 \pm 0.06$&$34.91 \pm 0.06$\\
TMTS J01380974+4723211& 59510.46057&$1.8 \pm 0.2$&$24043.14 \pm 221.77$&$802.58 \pm 219.88$&$5950.30 \pm 1630.21$&$2606.42 \pm 600.76$&&\\
TMTS J01442812+4327127& 59511.43865&$1.39 \pm 0.09$&$10111.79 \pm 30.33$&$308.13 \pm 32.77$&$2284.47 \pm 242.99$&$781.14 \pm 63.72$&$34.02 \pm 0.04$&$34.56 \pm 0.04$\\
TMTS J02012142+7430235& 59185.48165&$1.68 \pm 0.12$&$12768.77 \pm 44.10$&$398.37 \pm 44.40$&$2953.47 \pm 329.17$&$1217.49 \pm 116.26$&$33.86 \pm 0.04$&$34.39 \pm 0.04$\\
TMTS J02093570+4117071& 59516.42661&$0.59 \pm 0.03$&$10384.88 \pm 31.26$&$447.27 \pm 38.17$&$3316.06 \pm 282.96$&$483.24 \pm 28.44$&$34.52 \pm 0.03$&$35.06 \pm 0.03$\\
TMTS J02354528+3429199& 58856.43668&$0.31 \pm 0.02$&$14320.31 \pm 117.87$&$1257.02 \pm 124.48$&$9319.51 \pm 922.91$&$711.99 \pm 54.25$&$33.32 \pm 0.03$&$33.85 \pm 0.03$\\
TMTS J02382397+4342369& 59196.44294&$1.93 \pm 1.45$&$8203.26 \pm 235.19$&$152.68 \pm 73.42$&$6697.61 \pm 249.35$&$470.47 \pm 326.08$&$34.8 \pm 0.3$&$35.4 \pm 0.3$\\
TMTS J02420089+3852565& 59194.41678&$0.63 \pm 0.04$&$8191.15 \pm 48.05$&$551.26 \pm 51.83$&$4087.01 \pm 384.28$&$635.27 \pm 44.95$&$35.06 \pm 0.08$&$35.60 \pm 0.08$\\
TMTS J02461232+3409287& 58856.43675&$1.94 \pm 1.08$&$15514.72 \pm 110.41$&$133.16 \pm 104.63$&$1327.24 \pm 609.51$&$334.34 \pm 91.09$&$35.37 \pm 0.13$&$35.91 \pm 0.13$\\
TMTS J02515141+4434084& 59196.44292&$0.47 \pm 0.03$&$1445.18 \pm 16.24$&$195.78 \pm 19.91$&$1451.47 \pm 147.58$&$168.94 \pm 11.84$&&\\
TMTS J02552795+5815041& 59540.49373&$0.077 \pm 0.008$&$25921.24 \pm 105.83$&$661.30 \pm 127.24$&$4902.87 \pm 943.36$&$91.97 \pm 13.22$&$33.68 \pm 0.07$&$34.22 \pm 0.07$\\
TMTS J02570023+4642138& 59531.45494&$0.09 \pm 0.03$&$16576.65 \pm 78.11$&$181.30 \pm 88.47$&$1371.80 \pm 626.42$&$26.64 \pm 7.93$&$32.33 \pm 0.13$&$32.87 \pm 0.13$\\
TMTS J03060058+2807084& 59199.42082&$0.31 \pm 0.02$&$16021.51 \pm 31.46$&$269.06 \pm 36.33$&$1994.83 \pm 269.36$&$149.29 \pm 13.00$&&\\
TMTS J03100245+4346147& 59530.42781&$1.09 \pm 0.09$&$22892.86 \pm 52.47$&$421.81 \pm 52.98$&$3127.29 \pm 392.83$&$834.05 \pm 81.08$&$34.24 \pm 0.06$&$34.78 \pm 0.06$\\
TMTS J03111364+4526272& 59530.42608&$2.57 \pm 0.06$&$6641.61 \pm 16.01$&$507.30 \pm 18.05$&$3761.08 \pm 133.81$&$2379.90 \pm 54.71$&&\\
TMTS J03143182+5700585& 59540.49371&$1.8 \pm 0.3$&$30350.10 \pm 16.01$&$80.29 \pm 14.29$&$599.66 \pm 101.74$&$263.97 \pm 23.43$&$33.47 \pm 0.04$&$34.01 \pm 0.04$\\
TMTS J03232062+3459324$^{\star}$& 59542.44885&$1.8 \pm 0.7$&$520.98 \pm 91.39$&$110.03 \pm 93.57$&$1606.00 \pm 435.28$&$268.26 \pm 57.06$&&\\
TMTS J03402568+3501575& 59551.44639&$0.49 \pm 0.06$&$5690.88 \pm 26.06$&$161.11 \pm 32.31$&$1194.43 \pm 239.55$&$141.10 \pm 17.71$&$34.27 \pm 0.11$&$34.81 \pm 0.11$\\
TMTS J03474651+3635558& 59551.44639&$0.46 \pm 0.03$&$4967.16 \pm 69.10$&$576.78 \pm 75.67$&$4276.19 \pm 561.01$&$479.50 \pm 46.65$&$33.83 \pm 0.04$&$34.37 \pm 0.04$\\
TMTS J03475843+4142053& 59544.42431&$2.0 \pm 0.2$&$1149.86 \pm 11.18$&$165.05 \pm 18.24$&$1223.64 \pm 135.20$&$589.18 \pm 25.35$&$33.82 \pm 0.02$&$34.35 \pm 0.02$\\
TMTS J03525194+5121166& 59535.47132&$1.7 \pm 0.2$&$2680.23 \pm 22.36$&$118.28 \pm 21.24$&$877.86 \pm 155.55$&$355.15 \pm 51.57$&$33.22 \pm 0.06$&$33.76 \pm 0.06$\\
TMTS J03581344+5404505& 58879.44997&$0.34 \pm 0.03$&$2218.47 \pm 20.99$&$135.17 \pm 21.30$&$1002.11 \pm 157.92$&$84.02 \pm 11.07$&$33.32 \pm 0.06$&$33.86 \pm 0.06$\\
TMTS J04150931+1904271& 59215.47149&$0.268 \pm 0.012$&$1580.20 \pm 28.27$&$366.69 \pm 29.02$&$2718.62 \pm 215.13$&$179.31 \pm 10.96$&&\\
TMTS J04222375+4137066$^{\star}$& 59549.41282&$0.40 \pm 0.03$&$25057.97 \pm 36.75$&$371.49 \pm 39.56$&$2754.24 \pm 293.30$&$273.19 \pm 19.90$&&\\
TMTS J04242159+3706192& 59554.42546&$0.151 \pm 0.012$&$17701.39 \pm 23.04$&$160.99 \pm 21.24$&$1193.62 \pm 157.27$&$44.00 \pm 4.11$&$32.85 \pm 0.04$&$33.39 \pm 0.04$\\
TMTS J04254360+3501496& 59554.42448&$0.339 \pm 0.006$&$13597.77 \pm 16.93$&$540.64 \pm 17.59$&$4008.28 \pm 130.38$&$334.84 \pm 8.30$&$34.926 \pm 0.012$&$35.465 \pm 0.012$\\
TMTS J04282995+3747418& 59554.42546&$0.75 \pm 0.11$&$36050.83 \pm 41.10$&$196.85 \pm 41.98$&$1462.24 \pm 305.96$&$264.98 \pm 43.23$&$33.81 \pm 0.07$&$34.35 \pm 0.07$\\
TMTS J04291264+2154545$^{\star}$& 59543.52212&$0.12 \pm 0.03$&$21496.02 \pm 32.19$&$108.90 \pm 28.49$&$826.84 \pm 188.03$&$22.67 \pm 3.45$&&\\
TMTS J04460540+1553240& 59226.43493&$0.75 \pm 0.04$&$13714.27 \pm 17.03$&$190.44 \pm 20.10$&$1411.91 \pm 148.99$&$259.99 \pm 17.80$&$32.91 \pm 0.03$&$33.44 \pm 0.03$\\
TMTS J04514921+4341465& 59200.4207&$0.16 \pm 0.02$&$28507.51 \pm 82.29$&$442.61 \pm 85.90$&$3282.16 \pm 635.67$&$124.41 \pm 17.09$&$33.80 \pm 0.06$&$34.34 \pm 0.06$\\
TMTS J04541626+3459537& 59552.44263&$3.8 \pm 0.2$&$11162.34 \pm 27.40$&$234.64 \pm 25.87$&$1739.58 \pm 191.77$&$1607.87 \pm 140.08$&$34.06 \pm 0.06$&$34.60 \pm 0.06$\\
TMTS J04563049+4348567& 59200.4207&$0.88 \pm 0.12$&$13310.15 \pm 110.96$&$134.88 \pm 24.73$&$1885.63 \pm 358.93$&$213.56 \pm 20.02$&$33.83 \pm 0.04$&$34.37 \pm 0.04$\\
TMTS J00231410+3354241& 59149.44526&$3.1 \pm 0.2$&$3119.94 \pm 85.63$&...&$7807.72 \pm 659.86$&$3479.19 \pm 215.73$&$34.25 \pm 0.03$&$34.79 \pm 0.03$\\
TMTS J03070420+0352598& 59214.42278&$0.31 \pm 0.03$&$9490.83 \pm 13.37$&...&$1791.61 \pm 97.77$&$169.20 \pm 4.42$&$33.74 \pm 0.06$&$34.28 \pm 0.06$\\
TMTS J03264186+3253138& 59542.44193&$0.302 \pm 0.014$&$12276.96 \pm 245.48$&...&$3719.46 \pm 470.68$&$218.30 \pm 8.92$&$34.41 \pm 0.04$&$34.95 \pm 0.04$\\
TMTS J03324904+5755181& 59207.41269&$1.12 \pm 0.05$&$1580.75 \pm 14.68$&$246.78 \pm 16.01$&$1829.59 \pm 118.68$&$505.11 \pm 26.81$&&\\
TMTS J03324904+5755181& ...&$0.66 \pm 0.04$&$3881.77 \pm 27.02$&$275.15 \pm 27.55$&$2039.96 \pm 204.26$&$329.57 \pm 24.18$&&\\
TMTS J03392990+4117185& 59544.42431&$0.24 \pm 0.04$&$10771.05 \pm 26.16$&...&$3404.14 \pm 1562.95$&$148.79 \pm 18.71$&&\\
TMTS J04115590+1722272& 59216.4208&$0.192 \pm 0.012$&$18148.35 \pm 200.88$&...&$5173.48 \pm 1886.02$&$259.09 \pm 20.69$&$34.61 \pm 0.04$&$35.15 \pm 0.04$\\
TMTS J04192986+2145136& 59215.42573&$0.21 \pm 0.02$&$4461.18 \pm 19.31$&$148.80 \pm 20.86$&$2493.64 \pm 235.32$&$56.27 \pm 4.67$&$33.05 \pm 0.04$&$33.59 \pm 0.04$\\
TMTS J04192986+2145136& ...&$0.391 \pm 0.010$&$5896.90 \pm 17.36$&$425.38 \pm 18.43$&$3153.74 \pm 136.64$&$303.47 \pm 9.64$&$33.783 \pm 0.014$&$34.322 \pm 0.014$\\
TMTS J03211793+3714474& 59203.41734&$0.23 \pm 0.02$&$571.77 \pm 13.55$&$111.58 \pm 12.91$&$982.03 \pm 235.62$&$47.41 \pm 3.68$&$32.63 \pm 0.03$&$33.17 \pm 0.03$\\
TMTS J03211793+3714474& ...&$0.09 \pm 0.03$&$13013.41 \pm 31.36$&$83.56 \pm 34.64$&$888.33 \pm 178.81$&$12.98 \pm 3.55$&$32.07 \pm 0.12$&$32.61 \pm 0.12$\\
TMTS J03211793+3714474& ...&$0.23 \pm 0.02$&$16489.77 \pm 17.76$&$132.64 \pm 16.26$&$1492.15 \pm 391.81$&$55.22 \pm 4.12$&$32.70 \pm 0.03$&$33.24 \pm 0.03$\\
TMTS J03272410+7459407& 59567.43093&$0.24 \pm 0.02$&$19649.82 \pm 23.50$&$214.71 \pm 24.03$&$1940.26 \pm 490.57$&$93.07 \pm 6.48$&$34.18 \pm 0.03$&$34.72 \pm 0.03$\\
TMTS J03272410+7459407& ...&$0.11 \pm 0.02$&$30761.40 \pm 86.04$&$378.99 \pm 93.28$&$3718.36 \pm 807.12$&$75.04 \pm 11.31$&$34.09 \pm 0.07$&$34.62 \pm 0.07$\\
TMTS J03272410+7459407& ...&$0.067 \pm 0.009$&$37589.88 \pm 112.81$&$421.29 \pm 103.92$&$3862.87 \pm 816.34$&$51.04 \pm 11.30$&$33.92 \pm 0.10$&$34.46 \pm 0.10$\\
TMTS J03423772+3805512& 59551.44641&$8.16 \pm 1.65$&$-4.71 \pm 35.42$&...&$6475.52 \pm 36.12$&$1530.89 \pm 87.60$&&\\

		\hline
	\end{tabular}


	\begin{tablenotes}[normal, flushleft]
	        \footnotesize
	        \item
	\textit{Note}: Column (1): name of the TMTS flares; column (2): the time when  the TMTS observation started; column (3): the flux of the flare peak normalized to the quiescent state; column(4): the time of the flare beginning since the observation started; columns (5-8): the full-time width at half peak flux; the duration of the flare which is calculated by $t_{\rm stop}-t_0$, the integration from $t_{\rm stop}$ to $t_0$ occupy 0.9~ED; the equivalent duration; the logarithm of white-band energy; the logarithm of bolometric energy. 
	For multi-flare events, we only list the Start Time of the first flare, while for multi-peak events, we think they do not have the property $t_{1/2}$. $\star$ means the stars have unreliable Gaia cross-matched sources. This table is only a subset and the full table is in machine-readable form available online.
		\end{tablenotes}
	\end{threeparttable}
	
\end{table*}

\section{Flare properties}
The properties of flare stars can be quantified by several parameters, including peak flux, the equivalent duration and the released energy etc. In this section, we carefully examined the relations between these flare parameters ($E_{\rm L}$, ${\rm ED}$, $F_{\rm peak}$ and $t_{1/2}$), Gaia colors and spectroscopic parameters of flare stars. A tight $F_{\rm peak}$ $-$ ${\rm ED}$ relation is reported, and we perform a comparison with other flare samples. For those flare stars with the LAMOST spectra, we further study the possible effects of chromospheric activities on the resultant flares.

\label{re} 
\subsection{Gaia color}
\label{color}
\begin{figure*}
	\includegraphics[width=1.8\columnwidth]{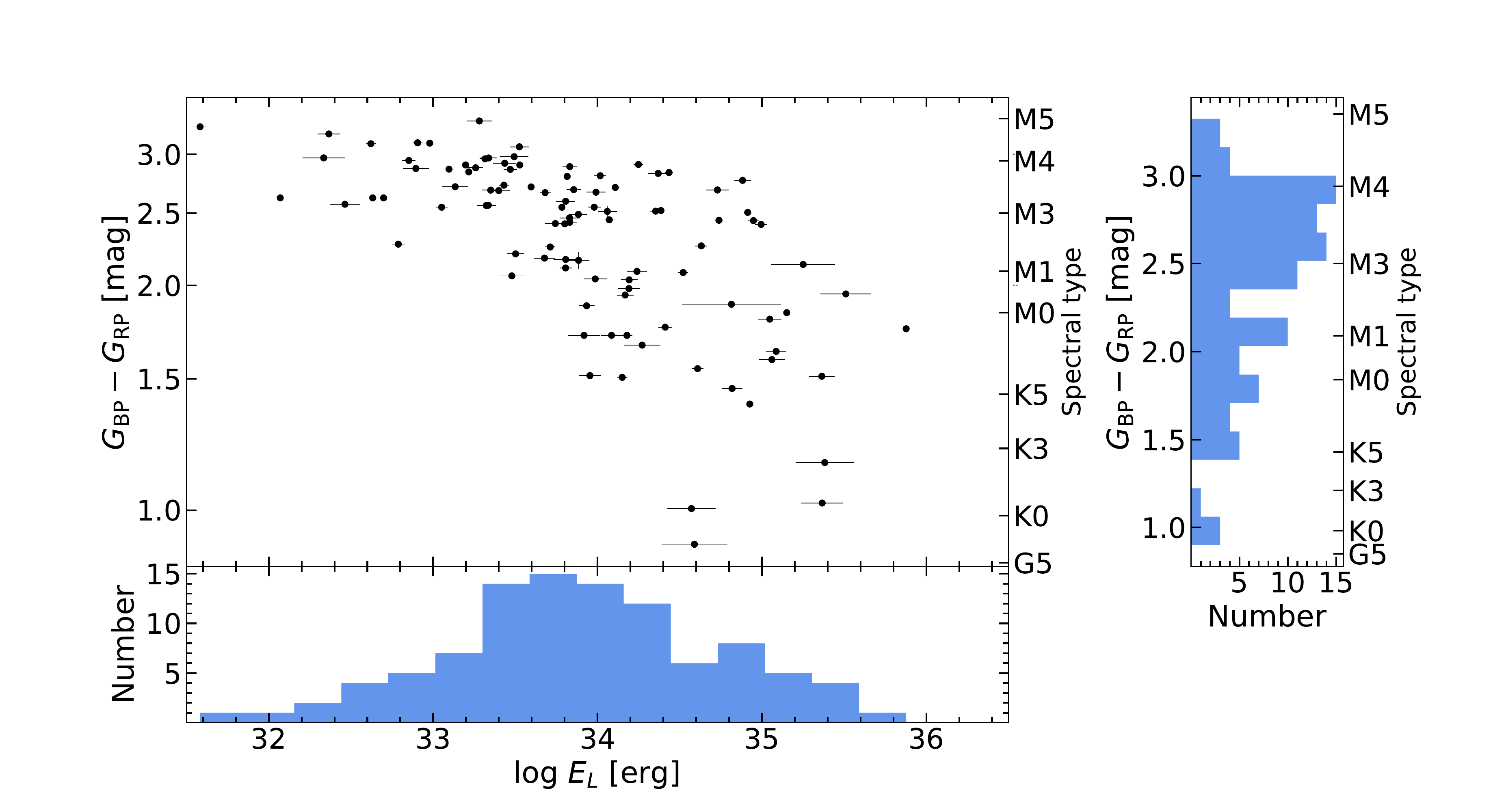}
    \caption{Distribution of the TMTS flare stars in the $G_{\rm BP}-G_{\rm RP}$ and  $E_{\rm L}$ space. The spectral type is given by \citet{2013ApJS..208....9P}.}
    \label{fig2}
\end{figure*}

\begin{figure}
	\includegraphics[width=\columnwidth]{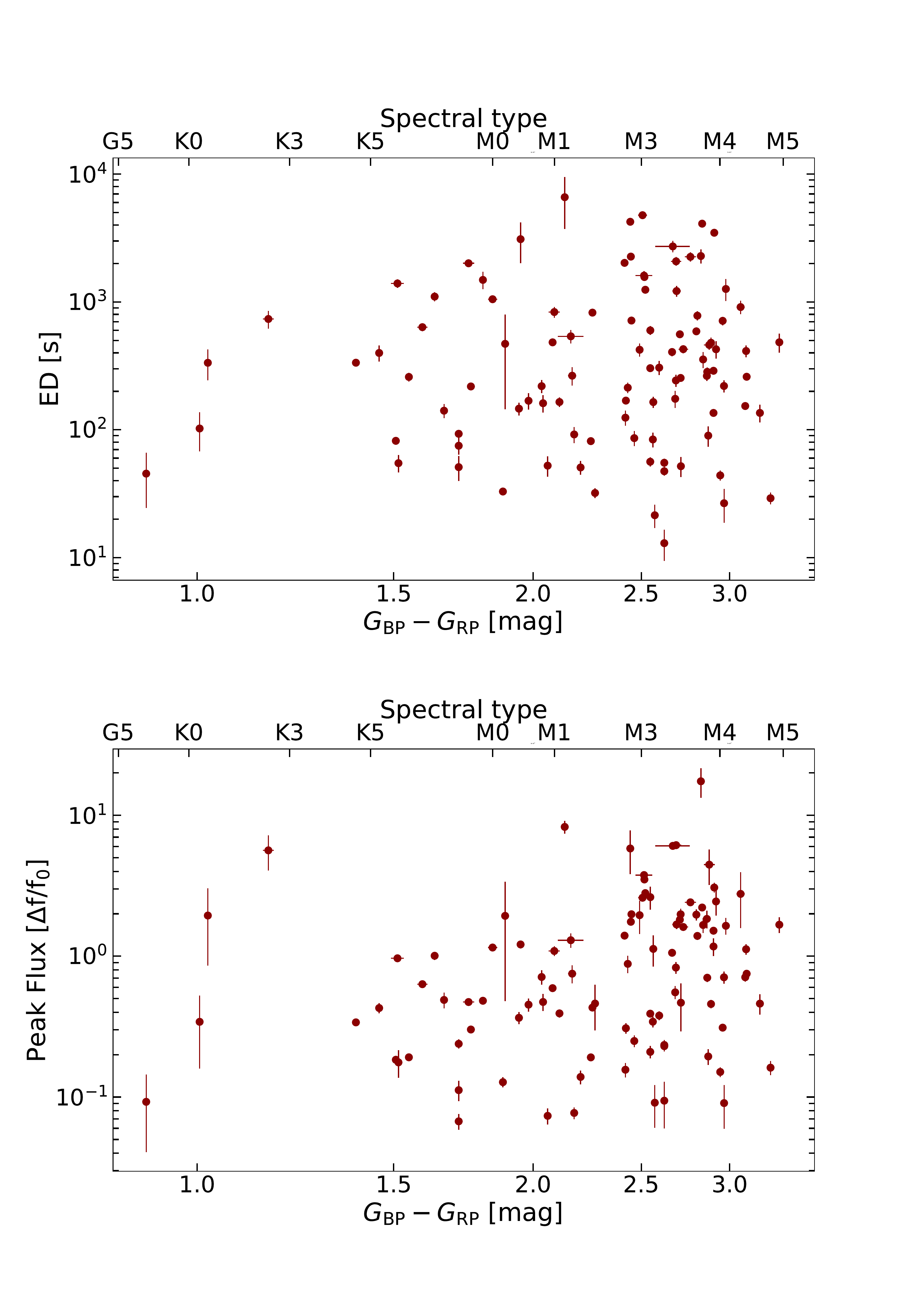}
    \caption{Relations between Gaia color ($G_{\rm BP}-G_{\rm RP}$), ED and peak flux of the flares. Top panel: $G_{\rm BP}-G_{\rm RP}$ versus ED. 
    Bottom panel: $G_{\rm BP}-G_{\rm RP}$ versus Peak flux. 
    }
    \label{fig3}
\end{figure}

 The relationship between peak flux, duration, and total energy of WLFs have been studied by previous works~\citep[]{2014ApJ...797..121H, Silverberg_2016}. Generally, they find that more energetic flares tend to have stronger peaks and longer durations. With the new TMTS sample, we can further explore the relations between flare properties and photometric/spectroscopic properties of the stars such as colors and spectral features. Fig.~\ref{fig2} shows the logarithm of the white-light energy of the flares (i.e., ${\rm log}~E_{\rm L}$) as a function of Gaia color of the stars (i.e., $G_{\rm BP}-G_{\rm RP}$). We used $G_{\rm BP-RP}$ to determine the spectral types shown in Fig.~\ref{fig2} and  Figs.~\ref{fig3} and \ref{t_half} as well, based on the intrinsic colour $-$ spectral type relation provided by \citet{2013ApJS..208....9P}\footnote{Here we have used the updated table in \url{https://www.pas.rochester.edu/~emamajek/EEM_dwarf_UBVIJHK_colors_Teff.txt} \label{spectral_estimate}}. By this way we can identify 75 M-type stars, 17 K-type stars and 1 G-type star in our sample. One can see that most of these flare stars have colors in the range of 2.5 mag $< G_{\rm BP}-G_{\rm RP} <$3.0 mag, which is consistent with the fact that flares from cooler dwarfs, especially those late K- and M-type dwarfs, are easily observed because of having high flare-occurrence frequencies and high variation contrast compared to their quiescent-state luminosity~\citep{Allred2015,  Muirhead_2018, 2019ApJ...881....9H}. Inspecting the TMTS sample, the flare energy does show a certain correlation with the Gaia color, with the spearman coefficient being as $r_s = -0.68$. In comparison with Kepler, TMTS detected fewer low-energetic flares, i.e., $E_{\rm L}$ $>$ 10$^{31}$ erg, which is likely due to the limited ability of distinguishing outburst signals from the noise of light curves (see discussions in section~\ref{ob}). This correlation between color and flare energy was previously studied, however, these studies focused only on maximum flare energy~\citep{2016ApJ...829...23D, 2020MNRAS.497..809J} and their sample is limited to cool stars ~\citep{2019ApJ...876..115S, Rodr_guez_Mart_nez_2020}. As shown in Fig.~\ref{fig2}, the average flare energy tends to be larger for hotter stars. However, this result might be affected by flare detection efficiency. In section ~\ref{efficiency}, we estimate the detection efficiency for our method and find that it will affect the energy distribution of flares. On the other hand, \citet{2021MNRAS.504.3246J} estimated the flare rates for different stellar types and they found that cooler stars follow a power-law fit with a roughly steeper slope compared to hotter ones, which is overall consistent with the observed trend of our flare sample. Nevertheless, more evidence or a larger and more complete sample is needed to verify this tendency in the future.
 

For the TMTS flare sample, we examine the relationships between the equivalent duration, the peak flux, and $G_{\rm BP}-G_{\rm RP}$ in Fig.~\ref{fig3}. As it can be seen, neither the relative peak flux (defined as $\Delta f/f_0$) nor the ED shows any significant dependence on the $G_{\rm BP}-G_{\rm RP}$ color. The spearman coefficient is $r_s=0.27$ ($p$-value 0.006) for peak flux and $r_s=0.06$ ($p$-value 0.52) for the ED, respectively. 



\begin{figure}
	\includegraphics[width=\columnwidth]{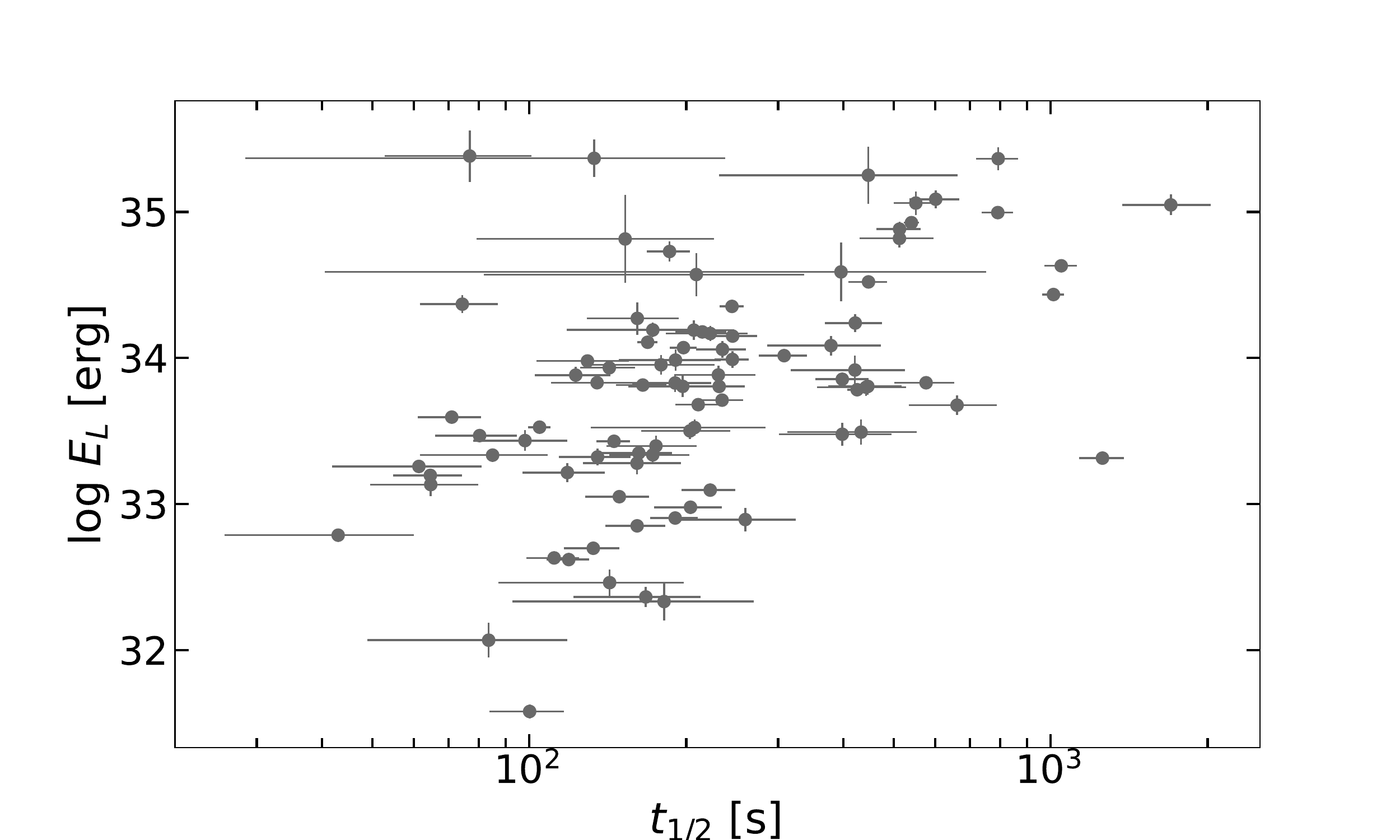}
    \caption{Relation between $t_{1/2}$ and $E_{\rm L}$. $t_{1/2}$ represents the full time width at half the maximum flux, calculated as $t_{\rm peak}-t_0$. $E_{\rm L}$ is the white-light flare energy.} 
    \label{fig4}
\end{figure}
\begin{figure}
	\includegraphics[width=\columnwidth]{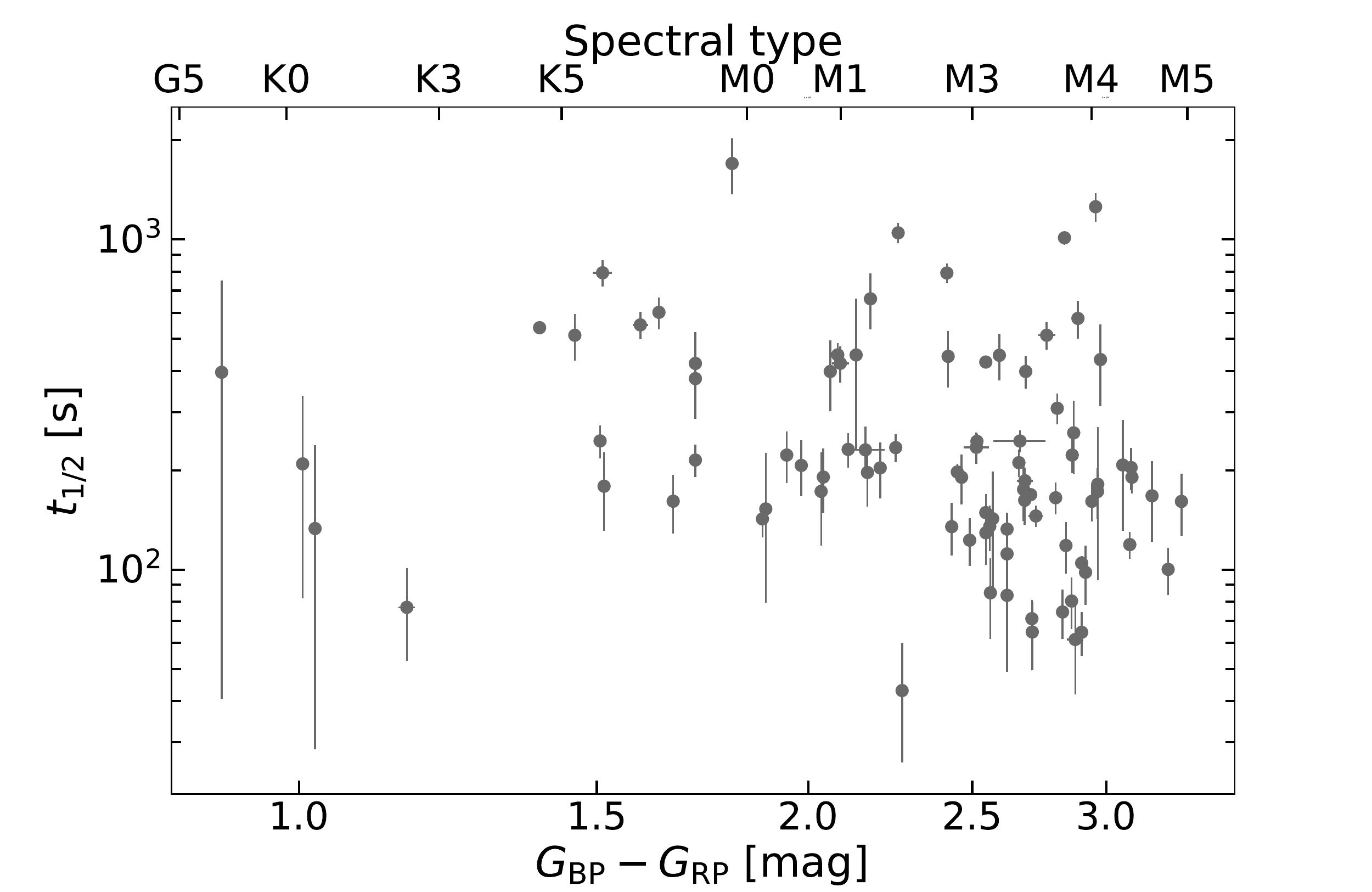}
    \caption{Relation between $t_{1/2}$ and ${\rm Gaia ~color}$ in our sample.} 
    \label{t_half}
\end{figure}

Furthermore, we use the classical flares in our sample (single-peak flare and multi-flare event) to examine whether the timescale $t_{1/2}$ has any correlation with the flare energy or the $G_{\rm BP}-G_{\rm RP}$ color. The relation between flare energy $E$ and duration $\tau$ has been reported in previous works. ~\citet{2015EP&S...67...59M} used Kepler data to study superflares on solar-type stars and found that the relation between the duration of the superflare $\tau$ and the flare energy $E$ follows $\tau \propto E^{0.39\pm0.03}$. A similar relation was also reported by ~\citet{2017ApJ...851...91N}, with $\tau \propto E^{0.38\pm 0.06}$ derived for the white-light flares detected in the Sun. The $E-\tau$ relation can be explained by magnetic reconnection, from which a theoretical scaling law $\tau \propto E^{1/3}$ has been derived~\citep[see,][]{2015EP&S...67...59M, 2017ApJ...851...91N}. From Fig.~\ref{fig4}, we get $t_{1/2} \propto E^{0.20\pm0.04}$. The deviation is caused by the measurement error, and we can get $t_{1/2} \propto E^{0.31\pm0.04}$ if considering the error bar of $E_{\rm L}$, which is roughly conformed with the results derived from magnetic reconnection. The corresponding spearman coefficient for the above correlation is $0.50$. Fig.~\ref{t_half} indicates that $t_{1/2}$ may show a weak dependence on the $G_{\rm BP}-G_{\rm RP}$ color, with the spearman coefficient $r_s$ being as $-$0.32 and the corresponding $p$ value being as 0.002, respectively. 
We caution, however, that the single timescale $t_{1/2}$ adopted in the model fit may not represent the best choice of describing the observed flare profile, which will be discussed in section ~\ref{Timescale}.



\subsection{The detection efficiency of our sample}
\label{efficiency}

As it can be seen from vertical histogram of Figure~\ref{fig2}, the flares occurring in late-time, cooler stars are more frequently detected. As flares with lower energy in hotter stars will have smaller amplitudes or $t_{1/2}$ values, it is naturally expected to detect fewer low-energy flares associated with early-type, hotter stars. This bias should be considered when discussing the statistical results in section ~\ref{color}.

\begin{figure}
	\includegraphics[width=\columnwidth]{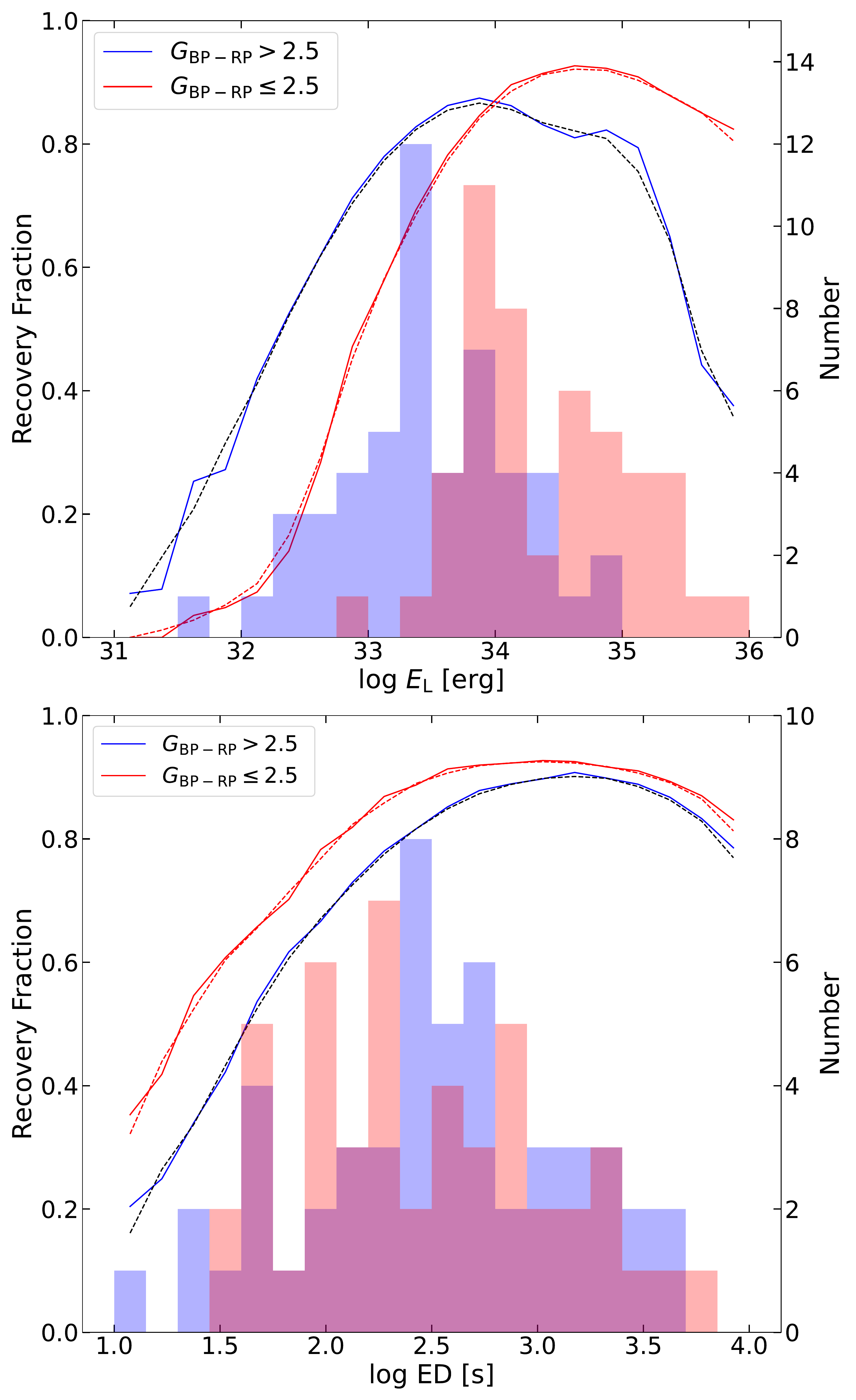}
    \caption{The recovery fraction as a function of energy and ED from the artificial flare injections of our sample. In the two plots, the blue line is for the subset with $G_{\rm BP-RP}>2.5$ mag, and red line is for $G_{\rm BP-RP}\leq2.5$ mag. The dashed lines represent the smoothed results, with the same corresponding colors. Top panel: The recovery fraction as a function of energy. The histograms indicate the energy distributions of the observed flares separated by $G_{\rm BP-RP}= 2.5 ~{\rm mag}$. Bottom panel: The recovery fraction as a function of ED. The histograms indicate the ED distributions of the observed flares separated by $G_{\rm BP-RP}= 2.5 ~{\rm mag}$.
    }
    \label{detection_efficiency}
\end{figure}

Our detection method for flares has been described in \citet{2022MNRAS.509.2362L}, where the light curve of each observation is identified as a flare candidate according to the estimated false discovery rate (FDR) of detrended flux. To test the efficiency of this method, we performed artificial flare injections and calculated the recovery fraction~\citep{2016ApJ...829...23D, 2020MNRAS.497..809J, 2020ApJ...905..107M}. For each flare star, the artificial flares are generated by our model in section~\ref{ob}, with $f_{\rm peak}$ chosen randomly within the range of 1.001-5 times $f_0$ (So $F_{\rm peak}$ is between 0.001 and 4). Here $f_0$ is the quiescent-state flux from our fitting results for each flare star in section~\ref{lightcurve} (See Eq.~\ref{eqmodel}). $t_{1/2}$ was chosen randomly between 30 s and 60 min. We used the observed time series and $f_0$ with $1 \sigma$ noise to simulate detrending light curves. We injected an artificial flare to the light curve and repeated 10000 times for every source. Then we calculated the FDR for each light curve, regarding a flare as being recovered if ${\rm FDR}<10^{-5}$. The recovery fractions were calculated as a function of TMTS $L$-band energy in 20 evenly-separated bins of log $E_{\rm L}$ = 31 to 36. We did not calculate the recovery fractions for each source. Instead, we divided our sample into two subsets, with $G_{\rm BP-RP}>2.5$ (Set 1) and $G_{\rm BP-RP}\leq2.5$ (Set 2), and calculated the recovery fraction just for each subset. Fig.~\ref{detection_efficiency} shows the result of our recovery tests. Following~\citet{2016ApJ...829...23D}, we applied a Weiner filter to smooth the recovery fractions. For the two cases (Set 1 and Set 2), we measured the minimum energy at which more than 68 percent artificial flares are recovered. The corresponding minimum energies log $E_{\rm min}$ are 32.875 and 33.375, respectively. We also plotted the distributions of observed flare energies for the stars in Set 1 and Set 2, where the detection efficiency in specific energy band seems to have an impact to the number of observed flares, and cause the shift of average energy from cool stars to hotter stars.

In the top panel of Fig.~\ref{detection_efficiency}, we can see clearly that the simulated detection efficiency of flares tends to decrease rapidly in cooler stars ($G_{\rm BP-RP}>2.5$) with higher energies. The reason is that higher energetic flares in cooler stars usually require larger $t_{1/2}$ comparable to the baseline, which will be likely missed by our detection method. We checked the $t_{1/2}$ of these non-recovery artificial flares in cooler stars with energy exceeding $10^{35}$ erg, finding that they all have larger values of $t_{1/2}$ (i.e., $> 0.4 {\rm h}$), and the median value is 0.844 h. This limits the capacity of TMTS to detect high-energy flares in cooler stars, which further smears out the correlation between color and flare energy discussed in section~\ref{color}.

We also check the detection efficiency as a function of ED, as shown in the bottom panel of Fig.~\ref{detection_efficiency}. We binned the calculated ED of artificial flares into 20 bins in logarithmic scale from $1$ s to $10^4$ s, and divided them into Set 1 ($G_{\rm BP-RP}>2.5$ mag) and Set 2 ($G_{\rm BP-RP}<2.5$ mag) as before. The functions of recovery fraction behave similarly in these two sets, and we did not see significant drop at larger ED. This is due to that, for flares with ED $\sim 10^4$ s, the corresponding values of $t_{1/2}$ are not as large as those with energy > $10^{35}$ erg in cooler stars. For instance, the median of $t_{1/2}$ of artificial flares with ED from $10^{3.5}$ s to $10^4$ s is 0.519 h in Set 1 and 0.518 h in Set 2, respectively. At smaller ED, the recovery fraction in Set 1 is smaller than that of Set 2, which may be caused by that fewer flares of low amplitudes are detected by TMTS in cooler stars.
We compare the peak flux ($\Delta f/f_0$) of recovered artificial flares with ED from 1s to 100 s for Set 1 and Set 2, and find that the former has a larger median of $\Delta f/f_0$ (i.e., 0.211 versus 0.177). Nevertheless, the detection efficiency cannot account for the discrepancy between cooler and hotter stars in $F_{\rm peak}-{\rm ED}$ relation, as discussed below (i.e. higher power law index in Set 1). Assuming the same detection efficiency for Set 1 and Set 2, more flares with low peak flux and low ED will be detected for cooler stars. As a result, the discrepancy in power law index will be even larger.


\subsection{Distribution of Peak flux and Equivalent duration}
\label{3.2}
To better understand the properties of the TMTS flare sample, we further examined the distribution of the measured peak flux and equivalent duration, and compare them with other flare samples such as those from the Kepler and Evryscope observations. The Kepler sample contains 3140 flares of 209 stars~\citep{2015MNRAS.447.2714B}, and the Evryscope sample includes 575 flare stars~\citep{2019ApJ...881....9H}. 
Both samples are obtained in white-light bands, they are thus suitable for the comparison. Note that the TESS sample also have 1228 flaring stars~\citep{2020AJ....159...60G}, but it is not included in the comparison for the lack of the ED parameter. The Kepler bandpass is very close to the TMTS $L$ band, which covers the wavelengths from 400 nm to 900 nm. We also include the flare sample from the Evryscope project for comparison, although it observes in $g^{\prime}$ bandpass (400 nm $\sim$ 600 nm). 



\begin{figure}
	\includegraphics[width=\columnwidth]{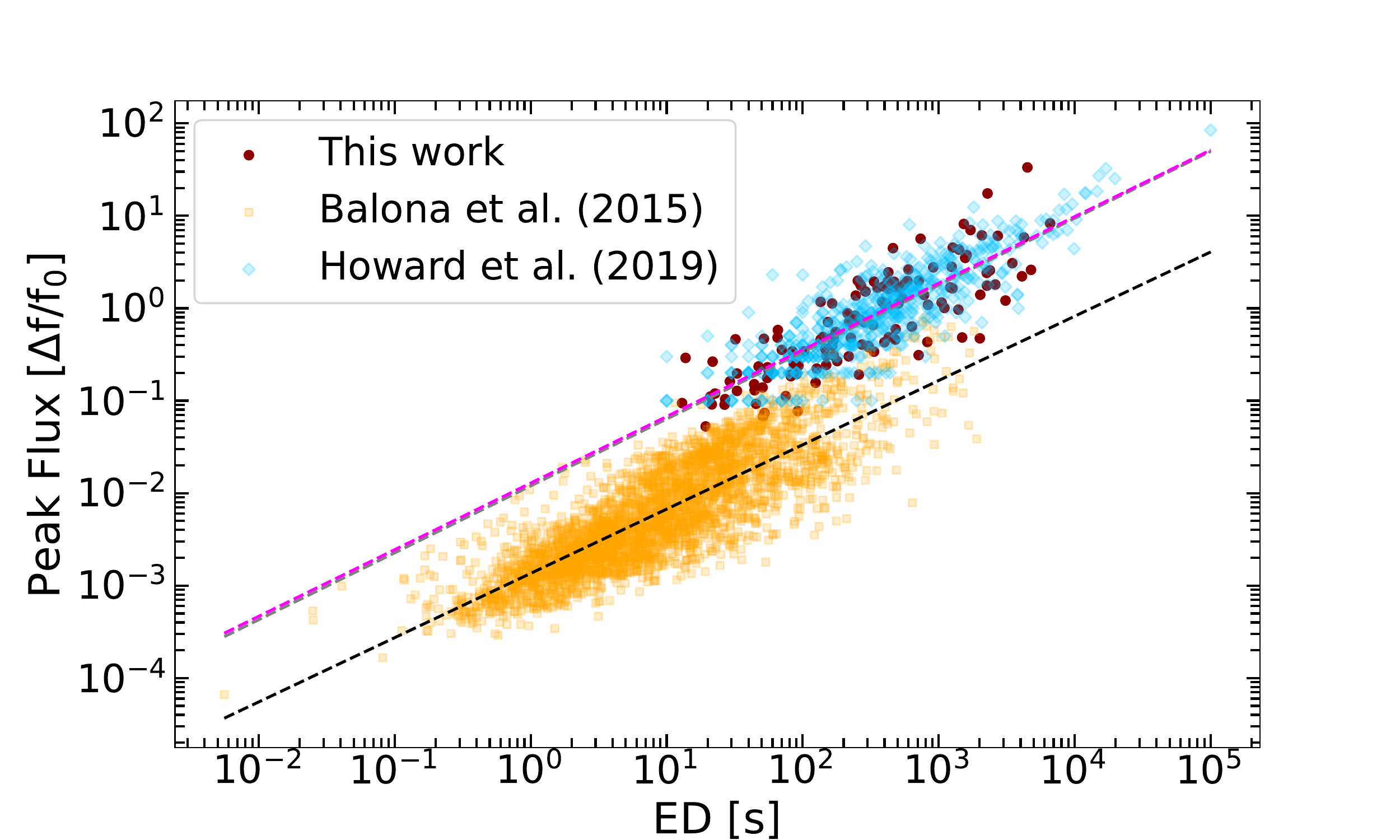}
	\caption{The equivalent duration versus peak flux. The orange squares and blue diamonds represent the data from ~\citet[][]{2015MNRAS.447.2714B} and ~\citet[][]{2019ApJ...881....9H}, respectively, while the red dots show the TMTS flares presented in this work. The dashed black, magenta, gray dash lines represent linear fits to the above three samples, respectively. The gray dash line is overlapped with the magenta line.}
	\label{fig7}
\end{figure}

Fig.~\ref{fig7} shows the peak flux and ED parameter measured for the flare samples from our sample, Kepler and Evryscope observations. As one can see these two quantities show a strong correlation for all of the above three samples, following a tight power-law relation. The best-fit relation yielded for the TMTS flare stars gives as $F_{\rm peak} \propto {\rm ED}^{0.72\pm 0.04}$, which is quite similar to that derived for the Evryscope sample (i.e., $F_{\rm peak} \propto {\rm ED}^{0.72\pm 0.02}$). This indicates that the properties of the TMTS flares show large overlapping with those of the Evryscope sample. The spearman correlation coefficients measured for these two samples are both about 0.84, with the corresponding \textit{p}-values being less than $10^{-4}$. The flare amplitudes of the Kepler sample tend to be lower than those of the TMTS and Evryscope samples at the same ${\rm ED}$, which is due to that the Kepler observations have higher base flux and this causes the translation of the linear fitting line in log $-$ log form. 

Since the Evryscope has a different bandpass (i.e., $g^{\prime}$), this result seems to show that the slope of $F_{\rm peak}-{\rm ED}$ relation is independent of bandpass. We further explored whether the bandpass has any effect on the slope.
Following the analysis of ~\citet{2013ApJS..209....5S} and ~\citet{Lin_2022}, we assume the flare emission as a blackbody, then we have the observed flare luminosity $L_{\rm flare}^{\prime}$ and the star luminosity $L_{\rm star}^{\prime}$ as
\begin{equation}
    L_{\rm flare}^{\prime}=A_{\rm flare}\int{B_{\lambda}(T_{\rm flare})R_{\lambda} d\lambda}.
	\label{eq:Lflare}
\end{equation}
\begin{equation}
    L_{\rm star}^{\prime}=\pi R_{\rm \star}^2 \int{B_{\rm \lambda}(T_{\rm eff})R_{\rm \lambda}d\lambda}.
	\label{eq:Lstar}
\end{equation}
Then the observed amplitude $F_{\rm peak}$ can be written as
\begin{equation}
    \frac{\Delta f}{f}= \frac{L_{\rm flare, obs}}{L_{\rm star,obs}}=\frac{A_{\rm flare}\int{B_{\lambda}(T_{\rm flare})R_{\lambda} d\lambda}}{\pi R_{\rm \star}^2 \int{B_{\rm \lambda}(T_{\rm eff})R_{\rm \lambda}d\lambda}} .
	\label{eq:amplitude}
\end{equation}
{$A_{\rm flare}$ is the area of flare events,  $B_{\rm \lambda}$ is Planck blackbody function, $R_{\lambda}$ is the response function, and $R_{\star}$ is the radius of the star. Recently, both spectroscopic and photometric studies showed that flares exhibit changing temperature between individual events~\citep[]{2013ApJS..209....5S, 2020ApJ...902..115H}.} According to ~\citet{2020ApJ...902..115H},  43\% flares are found to have a peak temperature larger than 14,000 K and 23\% are larger than 20,000 K. In our simulation, we assume the temperature of flare peak $T_{\rm eff, peak}$ is in the range from 8000 K to 20,000 K, and the effective temperature of star $T_{\rm eff}$ is from 3000 K to 5000 K. The $t_{1/2}$ parameter is set to vary from 2 min to 60 min. To simulate the TMTS flares and Evryscope flares, we perform random sampling for these three parameters for 5000 times, respectively. In the simulation, we first choose $A_{\rm flare}/{\pi R_{\rm \star}^2=0.01}$. With the randomly chosen $T_{\rm eff, peak}$ and $T_{\rm eff}$, we got $B_{\lambda}(T_{\rm flare})$ and $B_{\rm \lambda}(T_{\rm eff})$. Then we convoluted the two blackbody with transmission curve of Luminous filter and the quantum efficiency of the QHY 4040 CMOS detector ($g^{\prime}$ transmission curve), finding $\frac{\Delta f}{f_0}$ as in Eq.~\ref{eq:amplitude}. According to the light curve model shown in Eq.\ref{eqmodel}, we integrated the fractional form, $\Delta f/f_0$, to obtain the ED. Increasing the ${A_{\rm flare}}/{\pi R_{\rm \star}^2}$ ratio with a step of 0.005, we found the maximum amplitude of simulated data already achieve $10^2$ when ${A_{\rm flare}}/{\pi R_{\rm \star}^2}=0.015$, which exceeds the maximum value recorded in the observations (See Fig.\ref{fig10}). The simulated data of the TMTS simulated flares and the Evryscope flares show overlaps in the $F_{\rm peak}-{\rm ED}$ space, but the slopes derived for these two simulated datasets do have some differences (i.e., 0.66 vs. 0.74). We notice that the simulated data show larger scatter in the $F_{\rm peak}-{\rm ED}$ space relative to the observed ones (see Fig.\ref{fig7}). \citet{2020ApJ...902..115H} revealed that the peak temperature of flare emission has a dependence on the impulse. Moreover, Fig.~\ref{t_half} reveals a weak dependence of $t_{1/2}$ on the Gaia color. So the tight correlation of $F_{\rm peak}-{\rm ED}$ indicates $t_{1/2}$ shows some dependence on $T_{\rm eff, peak}$ or $T_{\rm eff}$ of the stars, which is ignored in our simulation.

\begin{figure*}
	\includegraphics[width=1.8\columnwidth]{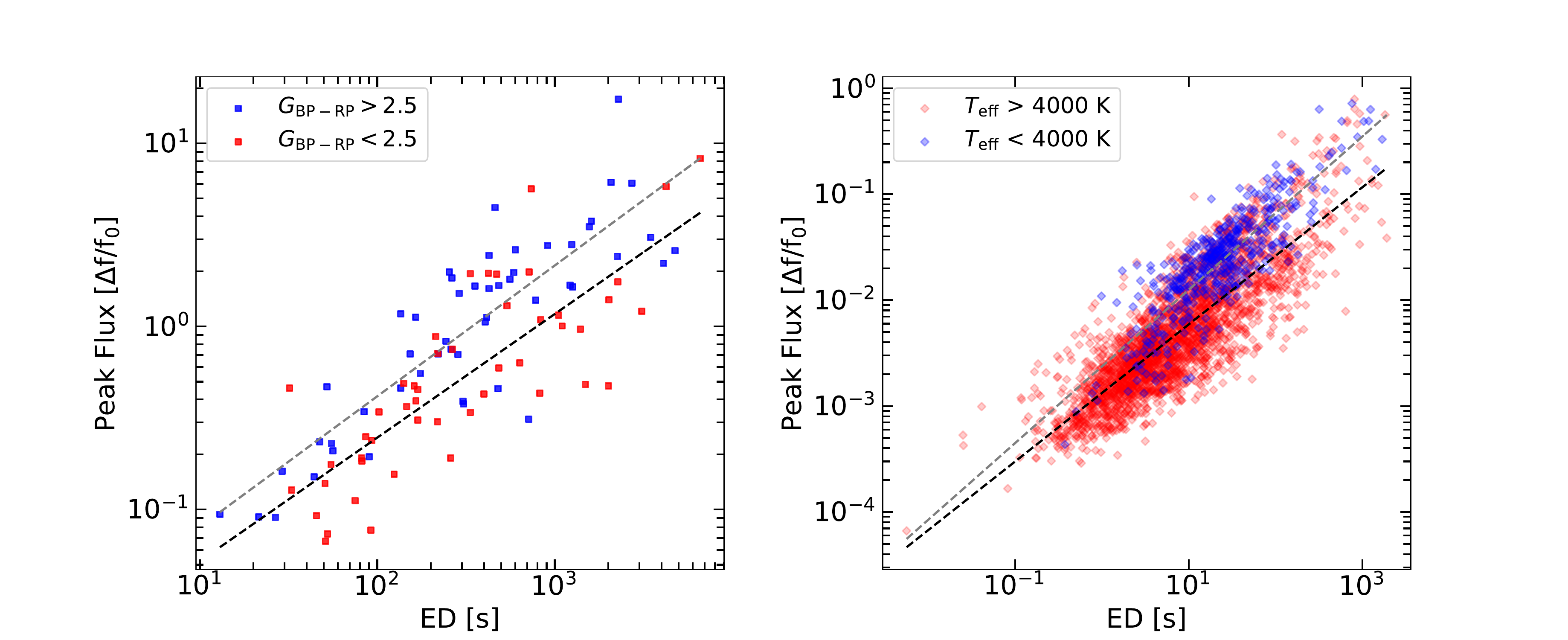}
	\caption{The same as in Fig.\ref{fig7}, but the samples are divided based on the Gaia color (TMTS sample) or $T_{\rm eff}$ (Kepler sample). Left panel: the TMTS sample, distinguished by blue squares for $G_{\rm BP-RP}< 2.5$ and the red squares for $G_{\rm BP-RP}>2.5$. Right panel: the Kepler sample, distinguished by blue diamonds for $T_{\rm eff} <4000 {\rm K}$ and red diamonds for $T_{\rm eff} > 4000 {\rm K}$. The grey and black dash lines represent the linear fits to the cooler and hotter stars, respectively.}
	\label{fig8}
\end{figure*}

\begin{figure*}
	\includegraphics[width=1.8\columnwidth]{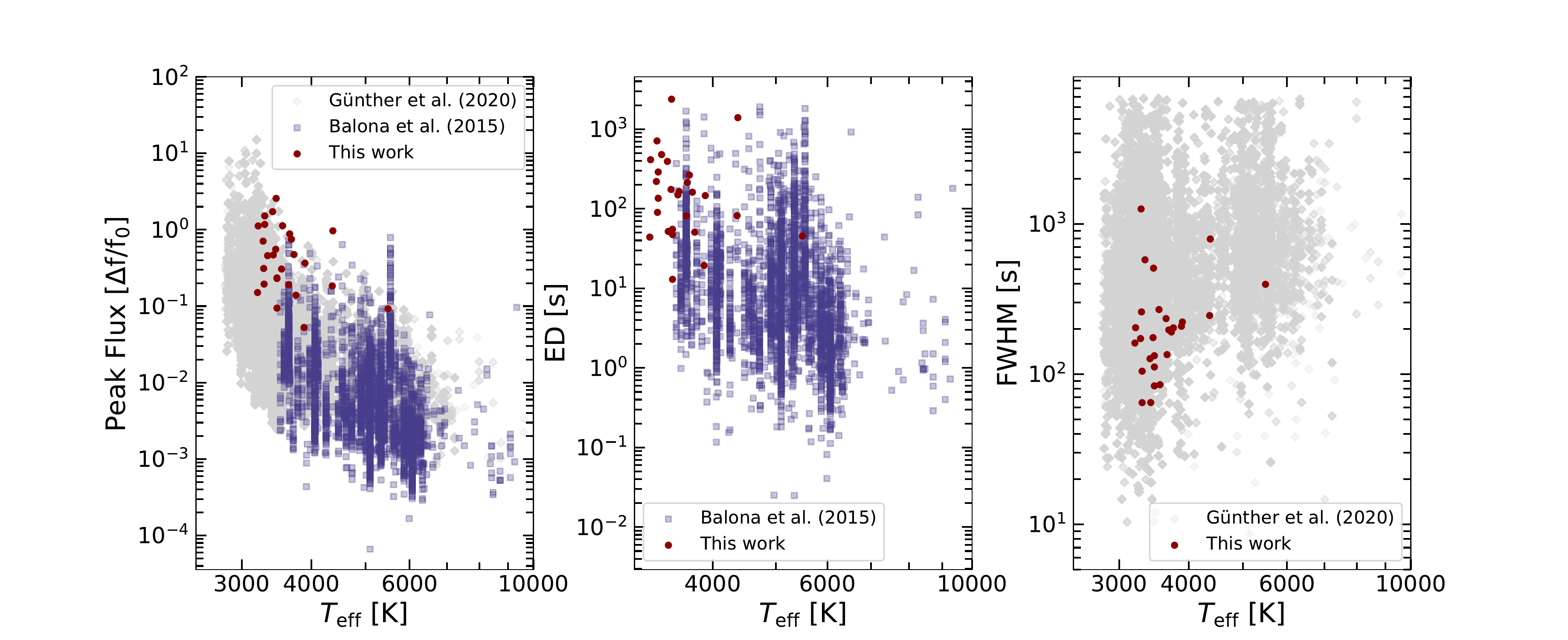}
	\caption{Correlations of peak flux and equivalent duration with the effective temperature $T_{\rm eff}$ for different flare samples. The purple squares represent the Kepler sample~\citep[][]{2015MNRAS.447.2714B}, while the grey diamonds represent the TESS sample~\citep[][]{2020AJ....159...60G}. The red dots illustrate the TMTS sample in our work. }
	\label{fig10}
\end{figure*}

To better explore the correlations shown in Fig.~\ref{fig7}, we divide the TMTS flare sample into two subgroups based on $G_{\rm BP-RP}$ = 2.5 mag, while the Kepler sample is also spit into two groups by effective temperature, i.e., $T_{\rm eff}$ = 4,000 K. The fitting results for different subgroups are shown in Fig.~\ref{fig8}. For the TMTS sample, a power law index 0.71 is obtained for the subgroup with $G_{\rm BP-RP}$ $>$ 2.5 mag, while an index of 0.67 is derived for the subgroup with $G_{\rm BP-RP}$ $<$ 2.5 mag. For the Kepler sample (see right panel of Fig.~\ref{fig8}), the power law indexes derived for hotter (with $T_{\rm eff}$ $>$ 4,000 K) and cooler stars (with $T_{\rm eff}$ $<$ 4,000 K) are $0.65$ and $0.72$, respectively. The spearman coefficients for the above correlations are all $>$0.8, and $p$-values are still less than $10^{-4}$. The separated fitting results show that for different subgroups, the slopes of linear fits are different, with hotter stars having higher values of ED even if their $F_{\rm peak}$ are the same as cooler stars. 
This trend shows that the shape of flare events do have a statistical change with the increase of $T_{\rm eff}$. In the case of having similar peak fluxes, larger ED means longer $t_{1/2}$ in our model. This result is consistent with the decreasing trend of $t_{1/2}$ as shown in Fig.~\ref{t_half}. 

In Fig.~\ref{fig10}, we further examine the correlations of peak flux and ED with the effective temperature of the flare stars. In addition to the TMTS sample, the samples from Kepler and TESS are overplotted for comparison. From the plot we can see that the relations of peak flux and equivalent duration with $T_{\rm eff}$ are not strong, 
which is similar to that described in section ~\ref{color}. The right panel of Fig.~\ref{fig10} shows the distribution of FWHM 
as a function of $T_{\rm eff}$, which indicates that flares of hot stars have relatively larger FWHM. This correlation is similarly revealed in Fig.~\ref{t_half}. 


\subsection{Spectroscopic Parameters}
\label{LAMOST}
In this subsection, we examine the spectroscopic properties of our flare sample. Since prominent ${\rm H \alpha}$ emission line in optical spectra of low-mass cool stars is proposed to form by collisional excitation in relatively dense chromospheres~\citep{West_2004,2018MNRAS.476..908F}, its strength (i.e., equivalent width EW$_{\rm H \alpha}$) can thus be used to trace the chromospheric activity and hence the magnetic field on the surface of the stars~\citep[e.g.,][]{West_2015, Newton_2017}.

\begin{figure}
	\includegraphics[width=\columnwidth]{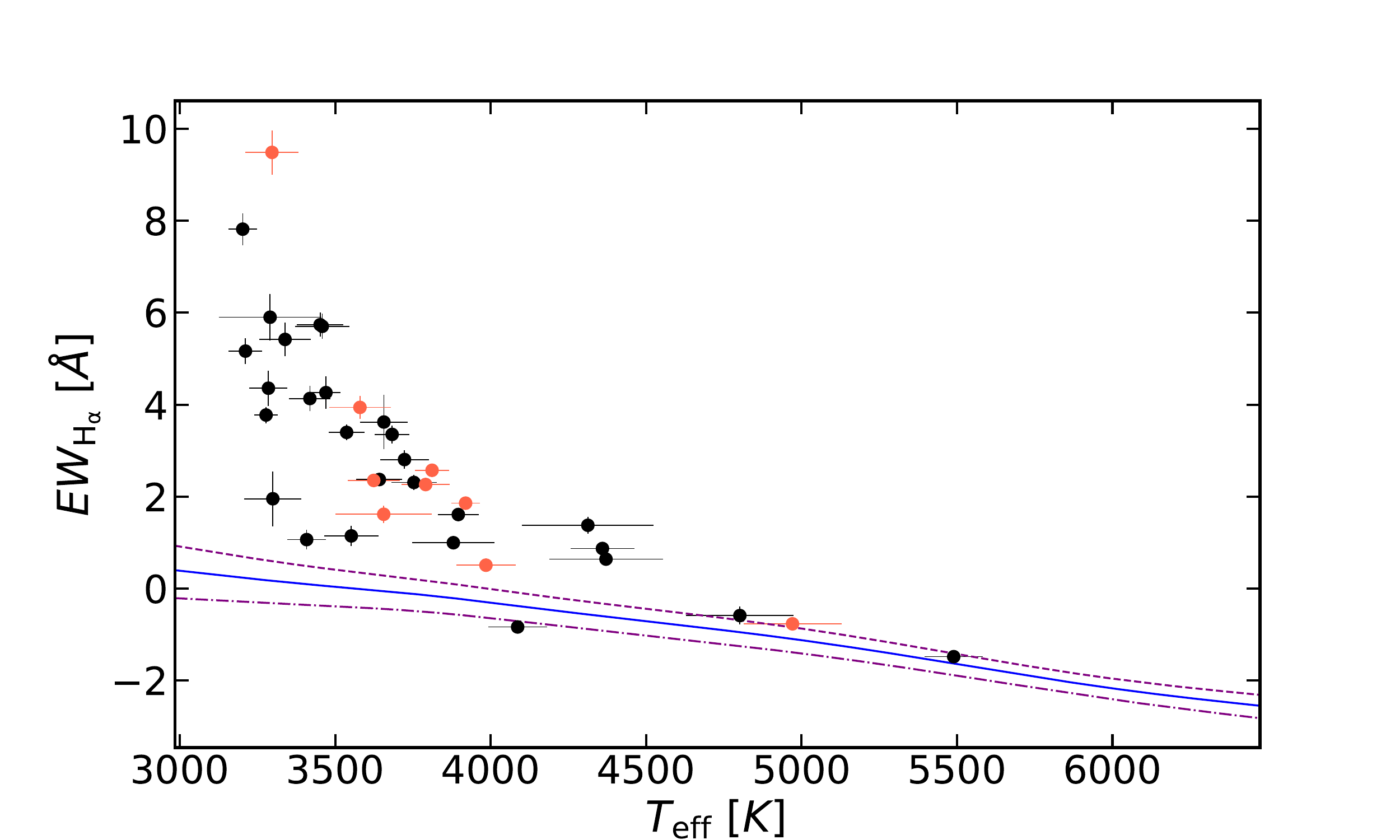}
	\caption{The relation of $EW_{\rm H_{\alpha}}$ and $T_{\rm eff}$. The blue line shows the mean value of ${\rm EW}_{\rm H_{\alpha}}$ derived from a larger sample of dwarf stars (~\citep{10.1093/mnras/stw1923}), with the purple dashed line and dot-dashed line representing the upper 3.0 $\sigma$ limit and lower 4.0 $\sigma$ limit, respectively. The black dots represent the classic events, while the red dots represent multi-peak events. All the data points correspond to individual stars. }
	\label{fig5}
	\includegraphics[width=\columnwidth]{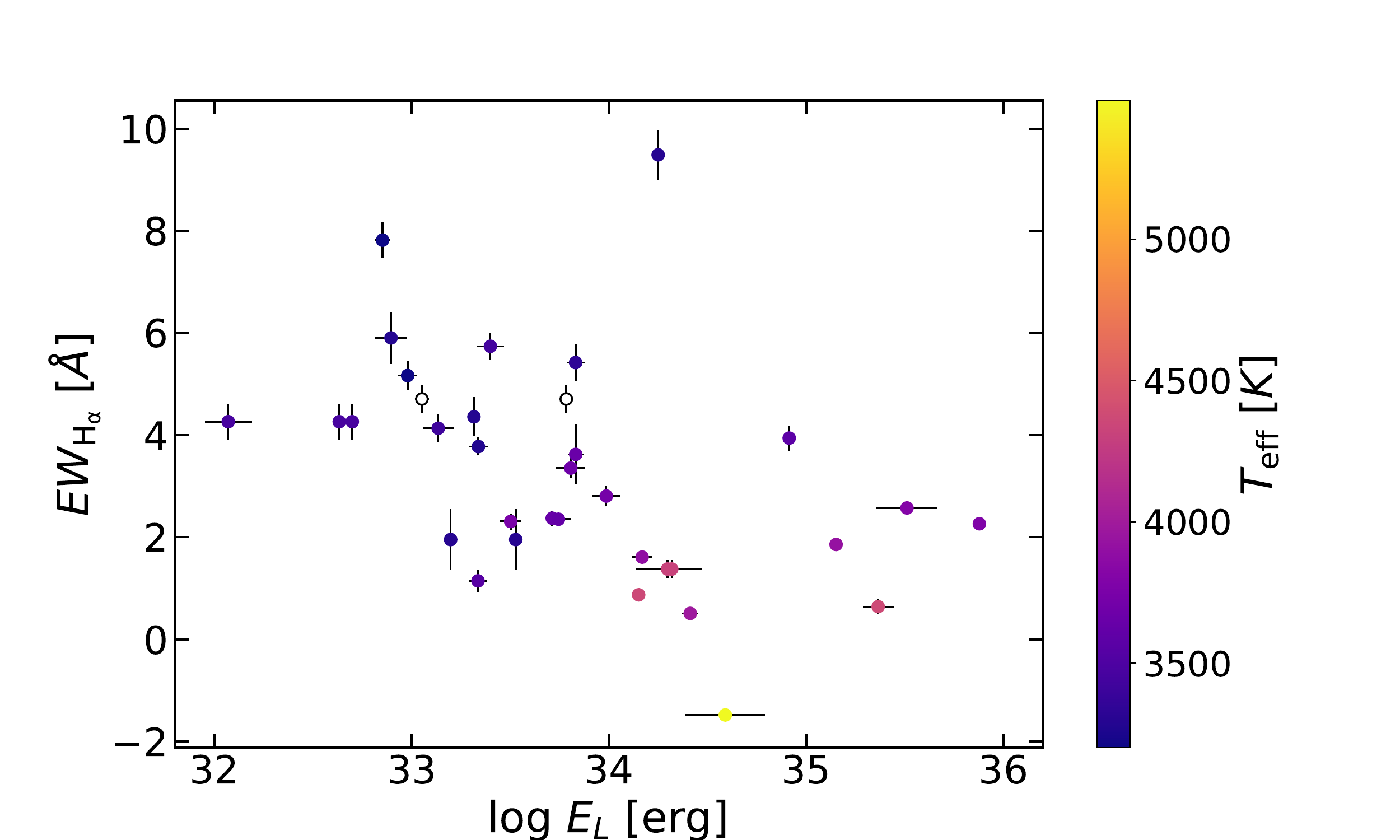}
    \caption{log $E_{\rm L}$ versus ${\rm EW}_{\rm H \alpha}$. The color gradient shows the effective temperature. The hollow dots indicate the stars without $T_{\rm eff}$ measurement.}
    \label{fig6}
\end{figure}

Using a large sample of dwarf stars with LAMOST spectra (N=16,4800), \citet{10.1093/mnras/stw1923} quantified the distribution of ${\rm H \alpha}$ emission as a function of effective temperature $T_{\rm eff}$ of these stars,  
and they found that the measured equivalent width of H$\alpha$ line (EW$_{\rm H\alpha}$) shows an anticorrelation with $T_{\rm eff}$, as shown in Fig.~\ref{fig5}. 
Following the definitions by \citet{10.1093/mnras/stw1923}, those with EW$_{\rm H\alpha}$ lying above the mean distribution by 3.5 $\sigma$ are identified as active stars, while those with EW$_{\rm H\alpha}$ lying below the mean value by 4.0 $\sigma$ are regarded as inactive ones. The EW$_{\rm H \alpha}$ and $T_{\rm eff}$ can be derived for 40 and 36 flare stars of our sample, respectively, and most of our flare sample located in the region of active stars (see Fig.~\ref{fig5}). The only two exceptions include a hot star TMTS J06032097+2136151 with $T_{\rm eff}$ = 5489 K and a cold star TMTS J05112829+5749355 with $T_{\rm eff}$ = 4087 K. As EW$_{\rm H \alpha}$ $-$ $T_{\rm eff}$ plot connects stellar activity to flare stars directly, it is naturally expected that cooler stars are more chromospheric-active. 
Note that all of the identified flare stars with $T_{\rm eff}\lesssim 4000~{\rm K}$ belong to chromospheric-active stars. With the increase in effective temperature, stars tend to become inactive. 
Previous investigations have better quantified the activity fraction of M dwarfs. For M0 stars, only a few percent are active, while this fraction increases to about 90\% for M7-M9 stars ~\citep{West_2008, Hilton_2010, 2019ApJ...876..115S}. Thus, it is not unexpected to see that most of our flare samples are active stars. It is clear that the observed ${\rm EW}_{\rm H \alpha}$ decreases with $T_{\rm eff}$, while the flare energy ($E_{\rm L}$) shows a contrary trend with the temperature as shown in Fig.~\ref{fig6}.

In Fig.~\ref{fig5}, we also highlight the subtype, multi-peak flares. To quantify the difference between them and typical flare events, we apply the two-dimensional Kolmogorov-Smirnov (2DKS)  test\footnote{\url{https://github.com/syrte/ndtest/blob/master/ndtest.py}}~\citep[]{1983MNRAS.202..615P, 1987MNRAS.225..155F} to the distributions of EW$_{\rm H\alpha}$ and $T_{\rm eff}$. The resultant $p$ value of 2D KS test is 0.43, suggesting that these two subtypes of our flare sample do not have any significant difference. This seems to imply that the chromosperic emission may be irrelevant with the occurrence of multi-peak flares. Nevertheless, the successive flares can be induced by sympathetic emission from separate nearby active regions or repetitive emission from the same active region~\citep{Davenport_2015, 2022ApJ...926..204H}, which may not leave detectable signatures in low-resolution, quiescent-state spectra.


\section{Discussion}
\label{discussion}

\subsection{\texorpdfstring{$F_{\rm peak}$}{TEXT} —— \texorpdfstring{${\rm ED}$}{TEXT} relation}
\label{peak_ed}

To our knowledge, the relation between peak flux and equivalent duration is seldomly investigated in current works. The reason is that $F_{\rm peak}$ $-$ ${\rm ED}$ relation is indeed a temporal nature of flare emission, and it may not reveal the information of heating mechanism. The common eruption scenario is likely related to the collision of non-thermal electron beams with plasma in footpoints of the loop, responsible for heating of the lower atmosphere. However, there are other possible mechanisms. For instance,  ~\citet{2018ApJ...859..143H} suggested the possible contribution of flare loops to the white-light emission, considering hydrogen recombination, hydrogen free–free process and Thomson scattering process. They calculated the flare amplitudes according to their theoretical physical processes and found a linear relation between amplitude and electron density in the log-log plot. The emission from flare loop can be dominant if the electron density $n_e$ of the loop is $>$ $10^{12}$ ${\rm cm^{-3}}$. Loop-prominent systems have been detected in several observations of solar flares ~\citep{Mart_nez_Oliveros_2013, Saint_Hilaire_2014, 2018ApJ...867..134J}.  

Following the theory described in ~\citet{2018ApJ...859..143H}, the flare amplitude can be expressed as 
\begin{equation}
    \frac{\Delta f}{f_0}= \frac{f-f_s}{f_s}=\frac{A_f}{A_s}\frac{I-I_{\rm bg}}{I_s-\frac{A_f}{A_s}(I_s-I_{\rm bg})} .
	\label{eq3}
\end{equation}

Where $A_s$ is the surface area of the star, $A_f$ is the area of the flare, $I_s$ is the stellar surface intensity. $I_{\rm bg}$ is assumed to be equal to $I_{\rm spot}$, which is a function of spot temperature $T_{\rm spot}$. Note that $I$ is affected by electron density $n_e$ and the temperature of the loop $T_{\rm loop}$. If we assume the loop temperature as $T_{\rm loop}=10000~{\rm K}$, then $F_{\rm peak}$ can be computed with given $T_{\rm eff}$, $T_{\rm spot}$ and $n_e$. From Fig.~\ref{fig10}, $F_{\rm peak}$ shows a weak dependence on $T_{\rm eff}$. 
On the other hand, the little change seen in the intercepts of Fig.2 and Fig.4 shown in ~\citet{2018ApJ...859..143H} suggests that the effect of $T_{\rm spot}$ be limited.
From the two plots we can also found $T_{\rm loop}$ only has a minor effect on $F_{\rm peak}$. 
So we can estimate the influences from $T_{\rm eff}$, $T_{\rm loop}$ and $T_{\rm spot}$ are the reason of the dispersion seen in Fig.~\ref{fig7}, since they are all weak.
If we ignore the effect of these factors, the linear $F_{\rm peak}-n_e$ relation in the log-log plot in \citet{2018ApJ...859..143H} implies that the ED correlates with $n_e$ as a power law. Nevertheless, we can not conclude that these results are incompatible with non-thermal electron beam heating model, as the relation actually depends on how the heating evolves with time. So the underlying physics governing the luminosity evolution during the flare eruptions is still uncertain.

We notice that, in Fig.~\ref{fig8}, the slope of the linear fit to the F$_{\rm peak}-$ED relation shows difference for subgroups of hotter and cooler stars, suggesting that the heating and cooling mechanism may be different for stars of these two groups.  
From the scaling law presented in ~\citet{2017ApJ...851...91N}, $\tau \propto E^{1/3}B^{-5/3}$, we can derive that $\tau$ increases with higher $T_{\rm eff}$ because of the increasing trend of $E$ and decreasing trend of $B$~\citep{2015MNRAS.447.2714B}. This  can explain the intercept differences in the best fit linear lines as shown in Fig.~\ref{fig8}. 

Based on the TMTS $L$-band filter and the quantum efficiency of detector ~\citep{2020PASP..132l5001Z}, we can get an observation window that efficiently covers the wavelength region from 400 nm to 800 nm. The spectral energy peak of the quiescent stars will shift to the shorter wavelength with increase of effective temperature. This may explain the correlation between $F_{\rm peak}$ and $T_{\rm eff}$. Based on the Gaia color-$T_{\rm eff}$ relation derived by  \citet{2021MNRAS.507.2684C}, we calculate $T_{\rm eff}$ of our sample by assuming solar metallicity ${\rm [Fe/H]} = 0$ and ${\rm log}~g=4$~\citep{2022MNRAS.514.5528L}. Assuming that the quiescent emission of each star is a blackbody emission of $T_{\rm eff}$, we create a 3000 K blackbody spectrum with the same bolometric luminosity.
The two blackbody spectra were then convoluted with the transmission curve of Luminous filter and quantum efficiency of the CMOS detector, from which we can get the ratio $R=f_{\rm L, 3000K}/f_{\rm L, T_{\rm eff}}$. We calculate $F_{\rm peak}=(f_{\rm peak}/Rf_0) -1$ by this ratio to eliminate the temperature-dependent effect of different energy proportion in $L$ band. In the end, the $F_{\rm peak}$- Gaia color correlation, with the spearman coefficient $r_s = 0.013$ and a $p$-value $0.90$, does seem to weaken in comparison with the correlation shown in Fig.~\ref{fig3}. Thus, the correlation between $F_{\rm peak}$- Gaia color can be explained by the change of energy proportion in observed band in quiescent-state related to temperature. However, it is not the case of the discrepancy in $F_{\rm peak}$-ED relation of cooler stars and hotter stars in Fig.~\ref{fig8}. We recalculate ED with quiescent-state flux as $Rf_0$ for classical flares, and find the cooler stars ($G_{\rm BP-RP}>2.5$ mag ) still has a higher power law index than the hotter stars ($G_{\rm BP-RP} <2.5$ mag ) (i.e., 0.86 versus 0.72).

Note that the adoption of 3000 K blackbody spectrum is just used to provide a standard $L$-band energy proportion, and the exact adopted temperature will not affect the the result of above analysis. For example, if we use a 4000 K blackbody spectrum, we get $r_s=0.14$ and $p$-value 0.21; while for a 5000 K blackbody spectrum we get $r_s=0.13$ and $p$-value 0.30.  The correlations become even more marginal compared to the original case ($r_s=0.27$, see section ~\ref{color}). However, 3000 K is better as it ensures $F_{\rm peak}=(f_{\rm peak}/Rf_0) -1$ is positive for the analysis in the logarithmic space.


\subsection{Possible connection between \texorpdfstring{${\rm H \alpha}$}{TEXT}  emission and stellar dynamo}
Based on the analysis of flare frequency distribution, \citet{2019ApJS..241...29Y} found that flare events from F- to M-type stars all follow a similar power law index. This indicates that the flares from different types of stars may have the same mechanism, i.e., magnetic reconnection. As our flare sample also include stars from F-type to M-type and the distribution of the measured ${\rm EW}_{\rm H \alpha}$ is continuous, as shown in Fig.~\ref{fig5}, which can be also used as an evidence favoring for a similar origin of various flares. 

Rotation has been proposed to account for stellar activity ~\citep{2003A&A...397..147P, 2007AcA....57..149K, 2018MNRAS.479.2351W}, and is also regarded as a proxy of stellar dynamo. The well-known picture for this relationship is that rapid rotators show saturated activities; and for slower rotators, their activities show a power law dependence on rotations. For stars with solar-like structures, the underlying dynamo is thought to be $\alpha$ $-$ $\Omega$ dynamo, in which the magnetic field is sheared in a tachocline layer that is caused by interaction between convective envelope and radiate core. The fact that the activity-rotation relation also holds for the full convective stars suggests that they have a turbulent dynamo or solar-like dynamo~\citep{2016Natur.535..526W, 2018MNRAS.479.2351W}. 

The saturated and unsaturated regions seen in the activity-rotation relations and the scatter in the relations of K-G stars can be explained by separating the main-sequence stars into I-sequence and C-sequence subgroups\citep{Barnes_2003, 2019ApJS..241...29Y}. These two groups have different stellar dynamos, with the I-sequence stars having slower rotations and C-sequence stars being younger and having faster rotations. In addition, the magnetic dynamos are also different: the C-sequence stars have decoupled the convective zone from the radiative core; while the above two zones are still coupled in the I-sequence stars, where the magnetic field is from the interior of the convection zone. 
The C-sequence stars will finally evolve into the I-sequence stars when the magnetic dynamo changes to the solar-like dynamo. 

Chromospheric activity is also found to relate with rotation~\citep{2012AJ....143...93R, West_2015, 2018A&A...614A..76J}. ~\citet{Newton_2017} studied a sample of mid-late M dwarfs, and they found a power law relation between ${\rm H \alpha}$ emission ($L_{\rm H \alpha}/L_{\rm bol}$) and Rossby number ($R_o=P_{\rm rot}/\tau$, $P_{\rm rot}$ is the rotation period, and $\tau$ is convective overturn timescale) with an index of $-1.7\pm 0.1$. However, the $L_{\rm H \alpha}/L_{\rm bol}$-$R_o$ relation also shows a saturation similar to the activity-rotation relation. The break point reported by ~\citet{Newton_2017} is near $R_o=0.2$, which is also similar to the break point $R_o = 0.13$ seen in the activity-rotation relation given by~\citet{Wright_2011}. The case of ${\rm Ca ~\uppercase\expandafter{\romannumeral2}}$ emission was reported in ~\citet{2017A&A...600A..13A}, with the $R_{\rm HK}'$ index ~\citep[$R_{\rm HK}'=R_{\rm HK}-R_{\rm phot}$, a fractional chromospheric luminosity in Ca ~\uppercase\expandafter{\romannumeral2} lines, see,][]{1984ApJ...279..763N}. 

To explain the variations in ${\rm H \alpha}$ emission, we suggest that the ${\rm H  \alpha}$ emission of C-sequence stars depends more on $T_{\rm eff}$ while that of the I-sequence stars depends more on rotation.  ~\citet{Yang_2017} studied 89 M-type flare stars with the LAMOST spectra, and they found that only 50 of them showed ${\rm H\alpha}$ emission. Note, however, that all of the stars with rotation period shorter than 10 days show prominent ${\rm H \alpha}$ emission. This favors that rotation (or stellar dynamo) also plays an important role in producing stronger ${\rm H \alpha}$ and stellar activity. 
As the analysis above, the behavior of ${\rm H \alpha}$ emission is the same as the flare activity ${L_{\rm flare}/L_{\rm bol}}$ in ~\citet{2019ApJS..241...29Y}, which means that the chromosperic emissions are correlated with flares. However, since most of our observations have short baseline, i.e., t $\lesssim 12$ hours, we can only find periods less than $6$ hours. Thus, it is not possible to detect rotation periods for our sample.

\subsection{Dependence of flare energy on effective temperature }
The analysis presented in Section \ref{re} indicates that more energetic flares tend to occur in stars with bluer colors (or higher temperatures), which is consistent with the observations that superflares occur preferentially in some G-type dwarf stars \citep{2013ApJS..209....5S, Okamoto_2021} and the dependence of flare energy on spectral type~\citep{2019ApJ...876..115S, Rodr_guez_Mart_nez_2020}.
For stronger flares detected in such solar-type stars, outburst energy can be explained by magnetic energy stored in starspots~\citep{Notsu_2019}. 
The strength of ${\rm H \alpha}$ tends to decrease with the increase of detected flare energy (see also Fig.~\ref{fig6}). For lower-mass M dwarfs, the magnetic fields are dominated by the poloidal component~\citep{2008MNRAS.390..567M, 2008MNRAS.390..545D, 10.1093/mnras/stv1925}, while for more massive solar-type stars, the toroidal component dominates in the magnetic field~\citep{2009ARA&A..47..333D}. ~\citet[]{2013MNRAS.433.2445A} has studied the equilibrium structure of the magnetic field, with the assumption of non-barotropic fluid. They constructed an axisymmetric magnetic field structure, finding that the energy stored in the toroidal component can be significantly larger than the poloidal component. It is worth noting that \citet[]{2013MNRAS.435L..43C} constructed an axisymmetric equilibrium configurations of the twisted-torus geometries, and they discovered that the total magnetic energy tends to increase with strength of the toroidal component. Although these investigations were conducted under relatively simple assumptions that may not be the reality of our sample, as the dynamo mechanism like Tayler–Spruit dynamo~\citep[]{2002A&A...381..923S} is the source of the large-scale magnetic fields, the increasing toroidal component in hotter stars still remains a possible explanation.

 
\subsection{The Timescale of the Model}
\label{Timescale}
Note that our model described in section~\ref{ob} has only one timescale, $t_{1/2}$, which is the same as \citet{2014ApJ...797..122D}. This model has been recently tested by ~\citet{2022ApJ...926..204H} with 20-s cadence observations of 243 classical flares on M dwarfs, proving the reliability of the template adopted in our analysis. Only 8.6\% of the classical flares were found to be better described by a Gaussian, and the rest can be well fit by the adopted template. However, the reason why a single timescale is adopted is due to that the 1-min cadence observation can not guarantee to capture the rapid rise phase of the flares. It is a strong assumption to adopt a single timescale $t_{1/2}$ which imposes the decay phase of classic flares declines to nearly $1/e$ from $t_{\rm peak}$ to $(t_{\rm peak} + t_{1/2})$. 
Thus, we try to use two timescales to describe the rise- and decay-phase evolution, separately. The $t'$ parameter in the model function discussed in section~\ref{ob} can be written as $t'=\frac{t-t_{\rm peak}}{\tau_{\rm decay}}$ for the decay phase. So when $t=t_{\rm peak}+\tau_{\rm decay}$, we approximately have $f\sim f_{\rm peak}/e$ (See Eq.~\ref{eqmodel} at $t'=1$). We regarded $\tau_{\rm rise}=t_{\rm peak}-t_0$ as the timescale of rise phase, so $t'$ remains the same as in the rise phase ($t'=\frac{t-t_{\rm peak}}{t_{\rm peak}-t_0}$).

\begin{figure*}
	\includegraphics[width=2.0\columnwidth]{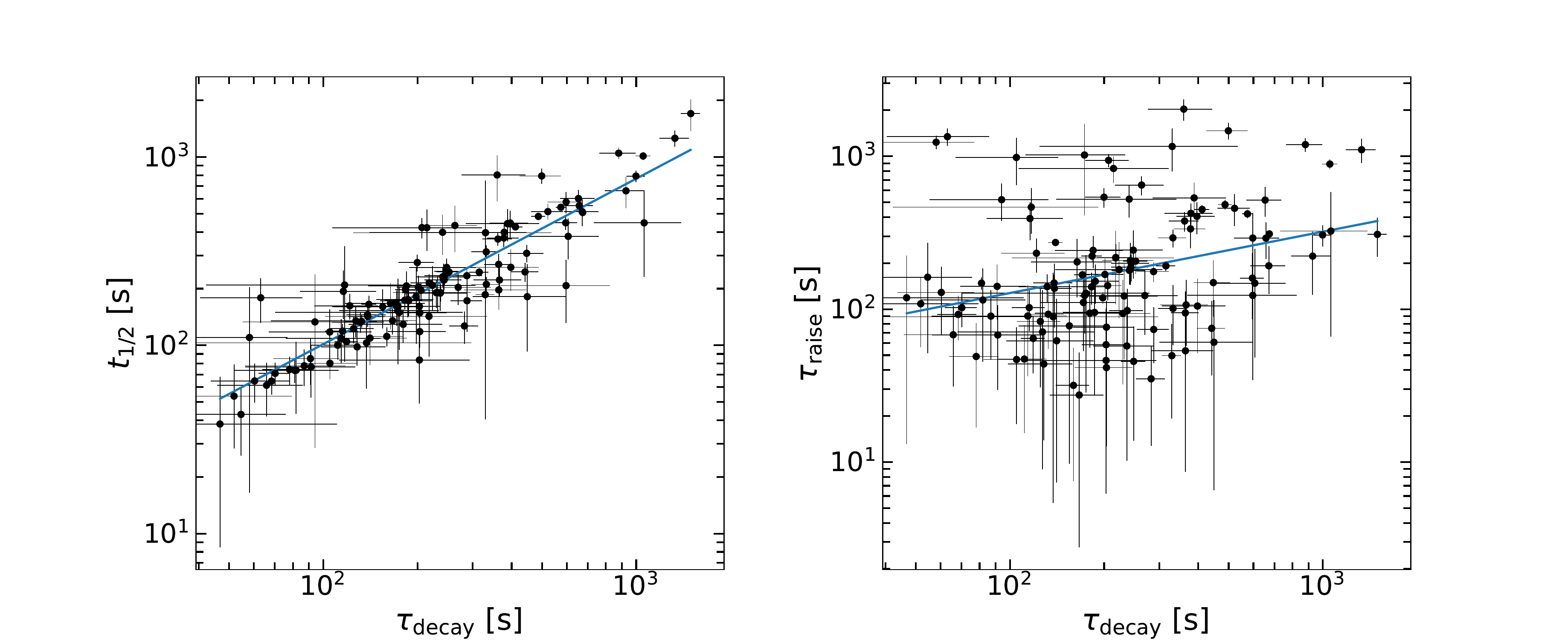}
    \caption{Left panel: the timescale $\tau_{\rm decay}$ in the two-timescale model, versus $t_{1/2}$ in the single-timescale model in section~\ref{ob}. Right panel: the  timescale $\tau_{\rm raise}$ plotted versus $\tau_{\rm decay}$ derived by two-timescale model. The blue lines represent the best linear fits to the data.}
    \label{fig9}
\end{figure*}

First, we compare $\tau_{\rm decay}$ with $t_{1/2}$, as shown in the left of Fig.~\ref{fig9}. These two timescales show a tight correlation, i.e., $t_{1/2}\propto {\tau_{\rm decay}}^{0.88 \pm 0.04}$, suggesting that the assumption of a single timescale is not strictly true. As $t_{1/2}$ is also the rise timescale in the single-timescale model, it is also possible that the rise phase is related to the decay phase. We used the two-timescale model to compare $\tau_{\rm rise}$ and $\tau_{\rm decay}$, getting $\tau_{\rm rise}\propto {\tau_{\rm decay}}^{0.4 \pm 0.1}$, as shown in the right panel of Fig.~\ref{fig9}. The spearman
coefficient is $r_s=0.34$, and corresponding $p$-value is $0.00011$. This result is roughly consistent with that yielded in ~\citet{2022PASJ...74.1069A}, supporting that the decay time of flares is dominated by the radiative cooling of the compressed chromosphere.
In the analysis of four giant flares from the TESS observations, \citet{10.1093/mnras/stab2159} set two parameters (${\rm FWHM_i}$ and ${\rm FWHM_g}$) to represent the impulsive phase and the general phase to decouple $t_{1/2}$. In our case, the purpose of adding $\tau_{\rm decay}$ is to use the decay-phase data to test the template, which is convenient to compare with $t_{1/2}$. However, the assumption of a single timescale still needs verification from observations with even shorter cadence in the future. 

\section{Conclusions}
\label{summary}
This paper presents observations and analysis of 125 flares from the first two-year survey of TMTS. We measured the properties of these flare events by an empirical template, and derived the parameters such as peak flux (normalized to the continuum) and the equivalent duration of the flares. With the distance and color parameters from Gaia DR2 and spectroscopic information from LAMOST DR7, we further explored the correlations between properties of flares and those of stars.  


We find a correlation between L-band flare energy $E_{\rm L}$ and Gaia color $G_{\rm BP}-G_{\rm RP}$. For hotter stars, the corresponding flares are on average more energetic compared with the cooler stars. Nevertheless, we also noticed that our detection efficiency shows some differences in cooler and hotter stars. For cooler stars, the limitation of observation duration goes against the detections of higher-energy flares. For hotter stars, it is difficult for the TMTS to capture lower-energy flares on them because of lower contrast of light variation. Thus, this trend still needs further investigations with larger and more complete flare sample. Equivalent duration (${\rm ED}$) of flare energy release is found to be independent of the $G_{\rm BP}-G_{\rm RP}$ color, while the ${\rm peak~flux}$ only shows a weak correlation with the Gaia color. The latter weak correlation is likely caused by change of energy proportion of the quiescent stellar spectra in the TMTS bandpass, as $T_{\rm eff}$ increases. 
The TMTS flare sample is also compared with flare samples from Kepler, TESS and Evryscope projects. The strong logarithm correlation found between $F_{\rm peak}$ and ${\rm ED}$, with a power law $F_{\rm peak} \propto {\rm ED}^{0.72\pm 0.04}$, is highly consistent with that derived from the Kepler and Evryscope samples. We also found a difference of the power law index between cooler stars and hotter stars. The correlation between $t_{1/2}$(the full-time width at half maximum flux) and $T_{\rm eff}$ (effective temperature of stars), and that between $t_{1/2}$ and $T_{\rm eff,peak}$ (temperature of flare peak) should play some roles in the tight relation between $F_{\rm peak}$ and ED. This correlation can also be explained by the contribution from flare loops, and it is expected that the equivalent duration is related to the electron density $n_e$ of the loop. 

 
To further explore the role of chromosphere in flare events, we measured the equivalent width of ${\rm H \alpha}$ line for 40 flare stars with spectra from LAMOST DR7. It is interesting to note that all the flare stars with lower effective temperature $T_{\rm eff}$ are chromospheric active, but hotter stars tend to be inactive. It is worth mentioning that the stars with multi-peak flares within a few hours do not show significant differences in distribution of EW$_{\rm H \alpha}$ and $T_{\rm eff}$. This suggests that chromospheric activity may not be the only factor triggering flare outbursts, which, however, should be further considered with a larger sample with spectral information (especially high-resolution spectra). The ${\rm H \alpha}$ emission in C-sequence stars is found to decrease with the increase of its temperature, while the I-sequence stars are found to have low ${\rm H \alpha}$ emission that is likely related to rotation. The different stellar dynamo may process magnetic field with different configurations, and low ${\rm H \alpha}$ emission should be expected if the magnetic field can not propagate deeply in the chromosphere (i.e., weak radial component). We also applied a two-timescale model to fit the flares and found $\tau_{\rm rise} \propto \tau_{\rm decay}^{0.4\pm0.1}$.   


In this work, we found several discrepancies in flare outbursts detected in hotter and cooler stars. For hotter stars, the released flare energy is larger, the index in $F_{\rm peak}-{\rm ED}$ relation is lower, and their flare profiles are broader. Analysis of our flare sample also indicates that, when the effective temperature increases, the chromospheric activity shows a decrease trend while the flare energy shows an increase trend, which can be explained by switch of stellar dynamo. All these clues hint that the change in effective temperature of a star would trigger differences of eruption physics. 

\section*{Acknowledgements}
We thank the anonymous referee for his/her suggestive comments which help improve this manuscript. This work is supported by the National Science Foundation of China (NSFC grants 12033003, 12288102, and 11633002), the Ma Huateng Foundation, the Scholar Program of Beijing Academy of Science and Technology (DZ:BS202002), and the Tencent Xplorer Prize. Y.-Z. Cai is funded by China Postdoctoral Science Foundation (grant 2021M691821). Qichun also thanks Xiang-Song Fang for his helpful suggestions on the measurement of ${\rm H \alpha}$ emission feature.  

This work includes the data from LAMOST (the Large Sky Area Multi-Object Fiber Spectroscopic Telescope) which is a National Major Scientific Project built by the Chinese Academy of Sciences. Funding for the project has been provided by the National Development and Reform Commission. 

We also use the data from the European Space Agency (ESA) mission Gaia (\url{https://www.cosmos.esa.int/gaia}), processed by the Gaia Data Processing and Analysis Consortium (DPAC, \url{https://www.cosmos.esa.int/web/gaia/ dpac/consortium}). Funding for the DPAC has been provided by national institutions,
in particular the institutions participating in the Gaia Multilateral Agreement.
\section*{Data Availability}

The data in Table.\ref{tab1} and Table.\ref{tab2} are available at the CDS. The light curves from two-year survey of TMTS will be available at TMTS Public Data Release  1\ (in prep.).



\bibliographystyle{mnras}
\bibliography{source} 








\bsp	
\label{lastpage}
\end{document}